\documentclass[fleqn,usenatbib]{mnras}

\usepackage{newtxtext,newtxmath}

\usepackage[T1]{fontenc}
\usepackage{ae,aecompl}
\usepackage{comment}

\usepackage{graphicx}	
\usepackage{amsmath}	
\usepackage{xcolor}
\usepackage[ruled,vlined]{algorithm2e}
\usepackage{soul}

\definecolor{temporary}{rgb}{0.7, 0.7, 0.01}

\definecolor{seagreen}{rgb}{0.190, 0.525, 0.361}
\definecolor{darksalmon}{rgb}{0.914, 0.588, 0.478}
\definecolor{steelblue}{rgb}{0.274 0.510 0.706}
\definecolor{operamauve}{rgb}{0.718 0.518 0.655}
\definecolor{firebrick}{rgb}{0.8 0.01 0.01}
\definecolor{antiquegold}{rgb}{0.8 0.543 0.0}

\title[Hierarchical generative models for star clusters]{ Hierarchical generative models for star clusters from hydro-dynamical simulations}
\author[Torniamenti et al.]{Stefano Torniamenti,$^{1,2,3}$\thanks{E-mail: stefano.torniamenti@studenti.unipd.it}
Mario Pasquato,$^{4,5}$
Pierfrancesco Di Cintio,$^{6,7,8}$ \newauthor
Alessandro Ballone,$^{1,2,3}$ 
Giuliano Iorio,$^{1,2}$
M. Celeste Artale,$^{9}$
Michela Mapelli$^{1,2,3}$\\
$^{1}$ Physics and Astronomy Department Galileo Galilei, University of Padova, Vicolo dell'Osservatorio 3, I--35122, Padova, Italy \\
$^{2}$ INFN- Sezione di Padova, Via Marzolo 8, I--35131 Padova, Italy \\
$^{3}$ INAF, Osservatorio Astronomico di Padova, vicolo dell'Osservatorio 5, 35122 Padova, Italy\\
$^{4}$ Département de Physique, Université de Montréal, Montreal, Quebec H3T 1J4, Canada\\
$^{5}$ Center for Astro, Particle and Planetary Physics (CAP$^3$), New York University Abu Dhabi\\
$^{6}$ Physics and Astronomy Department, University of Firenze, via G. Sansone 1, I--50019 Sesto Fiorentino, Italy\\
$^{7}$ INFN - Sezione di Firenze, via G. Sansone 1, I--50019 Sesto Fiorentino, Italy\\
$^{8}$ CREF, Via Panisperna 89A, I--00184 Rome, Italy\\
$^{9}$ Institut für Astro- und Teilchenphysik, Universität Innsbruck, Technikerstrasse 25/8, A-6020, Innsbruck, Österreich
}
\date{Accepted XXX. Received YYY; in original form ZZZ}
\pubyear{2020}
\begin{document}
\label{firstpage}
\pagerange{\pageref{firstpage}--\pageref{lastpage}}
\maketitle
\begin{abstract}
Star formation in molecular clouds is clumpy, hierarchically subclustered. Fractal structure also emerges in hydro-dynamical simulations of star-forming clouds. Simulating the formation of realistic star clusters with hydro-dynamical simulations is a computational challenge, considering that only the statistically averaged results of large batches of simulations are reliable, due to the chaotic nature of the gravitational $N$-body problem. While large sets of initial conditions for $N$-body runs can be produced by hydro-dynamical simulations of star formation, this is prohibitively expensive in terms of computational time. Here we address this issue by introducing a new technique for generating many sets of new initial conditions from a given set of star masses, positions and velocities from a hydro-dynamical simulation. We use hierarchical clustering in phase space to inform a tree representation of the spatial and kinematic relations between stars. This constitutes the basis for the random generation of new sets of stars which share the clustering structure of the original ones but have individually different masses, positions, and velocities. We apply this method to the output of a number of hydro-dynamical star-formation simulations, comparing the generated initial conditions to the original ones through a series of quantitative tests, including comparing mass and velocity distributions and fractal dimension. Finally, we evolve both the original and the generated star clusters using a direct $N$-body code, obtaining a qualitatively similar evolution.
\end{abstract}

\section{Introduction}

A large fraction of star formation happens in clusters (\citealt{2003ARA&A..41...57L}; but see also the recent findings by \citealt{2019MNRAS.486.5838R} and \citealt{2020MNRAS.495..663W}), which are
are clumpy and hierarchically sub-structured (\citealp{larson95,2009MNRAS.392..868B,2019A&A...629A.135D}). 
Young star clusters often show signatures of fractality (\citealp{Cartwright09,Kuhn19}), complex motions between sub-clumps (\citealp{Cantat-Gaudin19}), and possibly rotation (\citealp{Henault-Brunet12}). 
In addition, the early expulsion of gas due to stellar winds and supernova explosions brings young clusters out of dynamical equilibrium, causing an expansion phase (\citealp{hills80, goodwin06, Baumgardt07, pfalzner09}).
The comprehension of the early evolution of star clusters is of fundamental importance to interpret the later phases of their life. For example, \cite{2017NatAs...1E..64C} have shown that the rotation signature from the parent molecular cloud persists in the alignment of stellar spins in some open clusters, even today. 
Rotation in young and open star clusters is  confirmed  by numerical simulations \citep[e.g.,][]{2016A&A...591A..30L,2017MNRAS.467.3255M,2020MNRAS.496...49B}. Also, observations of globular clusters 
show signatures of rotation, sometimes with significant dynamical effects \citep[][]{2013ApJ...772...67B,2014ApJ...787L..26F, 2018ApJ...860...50F, 2018MNRAS.473.5591K, 2021MNRAS.506..813D} that may have been imprinted in the first phases of their evolution.

Gravitational $N-$body simulations are a key tool to model the early star cluster evolution, but they often start from rather idealized initial conditions, sampled from equilibrium models, such as \cite{1911MNRAS..71..460P} or \cite{1966AJ.....71...64K} models. Even though more sophisticated models are available \citep[e.g.,][]{1962MNRAS.123..447L, 1963MNRAS.126..269M, 1966AJ.....71...64K, 1970AJ.....75..674P, 1975AJ.....80..175W, 1984A&A...137...26B, 1987AJ.....93.1106L, 2005A&A...429..161T, 2006AJ....131..782A, 2012A&A...540A..94V, 2015MNRAS.454..576G, 2017MNRAS.468.1453D, 2019MNRAS.487..147C}, these were developed with the goal of describing the current, quasi-equilibrium state of star clusters, and do not perform well in describing the early and out-of-equilibrium structure of the clusters. Thus, by design, they bear little resemblance to observed primordial conditions in embedded clusters. 
For more than a decade, fractal initial conditions have been used as a starting point for realistic simulations \citep[e.g.,][]{2004A&A...413..929G, 2010MNRAS.407.1098A, 2011ascl.soft07015K, 2011MNRAS.417.2300K, 2011MNRAS.418.2565P, 2014MNRAS.438..620P, 2018MNRAS.478..183P, 2019MNRAS.487.2947D}, but even this approach does not guarantee that all the relevant characteristics of the actual primordial conditions of star clusters are correctly captured. 
 
A seemingly obvious but computationally demanding way to generate realistic initial conditions for star clusters is to run suites of hydro-dynamical simulations, coupled with appropriate recipes for handling star formation and other sub-grid physics \citep[e.g.,][]{2000ApJS..128..287K,2003MNRAS.343..413B,Bate09, 2012ApJ...761..156F, Krumholz12, Dale15, 2016ApJ...817....4F, Geen16, Seifried17, Zamora-Aviles19, Lee19, Wall19}. Despite these efforts, large sets of simulations including all the relevant physics are at present hard to come by, notwithstanding ever-advancing hardware capabilities.  This is compounded by the fact that $N-$body simulations, even direct-summation ones, eventually diverge from the true solution of the $N-$body problem for most initial conditions due to numerical errors and the chaotic nature of the problem (e.g. see \citealt{1993ApJ...415..715G, 2002ApJ...580..606H, 2003ApJ...585..244K, 2015ComAC...2....2B, 2019MNRAS.489.5876D, 2020IAUS..351..426D, 2020MNRAS.497.3694M, 2021arXiv210410843W} and references therein), with the consequence that only ensemble-averaged results are considered reliable within the current consensus. To obtain such averages, multiple $N-$body runs are needed, each with its own initial conditions.
 
The aim of this work is to introduce a novel approach to obtain initial conditions for $N-$body simulations, at a tiny fraction of the computational cost of running additional independent hydro-dynamical simulations. Our approach relies on a clustering algorithm, that is a method to identify similar instances in a sample and assign them to groups, or clusters. 
The usage of clustering algorithms is  widespread in cosmology, where different methods have been developed to identify over-dense gravitationally bound systems (i.e., groups or dark matter haloes and subhaloes). 
There is a huge variety of group-finding algorithms, which perform clustering in different ways.
For example, 
density-peak locators and spherical over-density finders identify density peaks in the particle distribution and draw spheres of decreasing density around them, down to a density threshold \citep[see, e.g.,][]{PressSchechter1974}. Direct particle collectors, instead, connect particles based on a linking length criterion. Such scheme is the basis of the friend-of-friends prescription, which is used in the context of halo-finding in cosmological simulations \citep[][]{Davis1985} and in galaxy clusters from observational data (\citealp{2012MNRAS.420.1861M}; see also \citealp{2017A&C....20...44F} and references therein). Other group finders extend these two approaches to include particle velocity information  \citep[see, e.g.,][]{Diemand2006,Maciejewski2009}.  

Here, we 
adopt a hierarchical clustering algorithm. As opposed to the aforementioned algorithms, hierarchical clustering does not rely on the definition of a density threshold or a length scale, but is provided with a definition of similarity between groups.
In particular, it proceeds in a hierarchical way, by connecting the most similar pair of clusters, starting from individual instances, until a certain number of groups is reached.
This hierarchical construction allows not only to identify over-densities in the distribution of stars, but also to  draw information about the structure of star systems at different scales. 
New stellar clusters can thus be obtained by modifying selected nodes in the hierarchical structure, depending on the properties we want to preserve or modify.

Our generative model is meant to produce new large-scale distributions of sink particles, by preserving the properties that make them so realistic, such as their complex fractal structure.
The goodness of our method is evaluated by comparing the fractal structure of the new realizations to that of the original cluster. Also, we  check if the new velocity distributions are consistent with the original distribution. 
Finally, we  run $N$-body simulations of the newly generated clusters to test if they are consistent with the stochastic fluctuations of the original simulations.



The paper is organized as follows. In Section~\ref{simu}, we recap the properties of the hydro-dynamical simulations we used to generate our original initial condition sets; Section~\ref{meto} presents our approach for generating new realizations, while in Section~\ref{resu} we describe our results and run various checks to compare the generated realizations to the original simulations. In Section~\ref{disc}, we discuss and draw conclusions.

\section{Smoothed-particle hydro-dynamical simulations}
\label{simu}

\subsection{Initial conditions and simulation set-up} \label{sec_set_up}

As a starting point for this work, we used the sink particles from  $10$ smoothed-particle hydro-dynamics (herafter SPH) simulations of molecular clouds performed by \cite{2020MNRAS.496...49B} using the {\sc gasoline2} code \citep[][]{2004NewA....9..137W, 2017MNRAS.471.2357W}. In the following, we may refer to these sink particles as `stars' for convenience.

The initial conditions of the SPH simulations are spherical molecular clouds with total gaseous mass in the range $10^4\leq M_{\rm mc}/{\rm M}_\odot\leq 10^5$  (see the last column of Table~\ref{table:alpha}), uniform temperature $T_0=10$~K and uniform density $\rho_0 = 2.5 \times 10^2$~cm$^{-3}$. All runs have a fixed number of initial SPH particles equal to $10^{7}$, corresponding to a gas mass resolution of $10^{-3}$ to $10^{-2}$ M$_{\odot}$, depending on the cloud mass. Stars  form during the simulation by means of a sink particle algorithm based on the same prescriptions as \citet{Bate95}. The spatial resolution (kernel size) of the simulation can get as small as 0.001 pc in the densest regions.

In order to induce a non-isotropic evolution, the SPH gas particles are initially given a turbulent, divergence-free, Gaussian random velocity field with a different random seed for each simulation of the set, following a \citet{Burgers48} velocity power-law spectrum with index $-4$ \citep{bate09b}. With respect to the classical \cite{kolmogorov41} power spectrum (with index  $-11/3$), the Burgers power spectrum better matches turbulence in compressive flows, where shocks are present (\citealp{federrath13}). The clouds are in an initial marginally bound state, so that their initial virial ratio $\alpha_{\rm vir}\equiv 2K/|W|=2$, where $K$ and $W$ are the gas kinetic and potential energy, respectively.

During the hydro-dynamical simulation, the gas equation of state has been set to be adiabatic, while radiative cooling by dust has been modeled as in \citet{Boley09} and \citet{Boley10}. The amount of energy lost by cooling was calculated through the divergence of the heat flux 
\begin{equation}\label{eflux}
\nabla \cdot F_{\rm cool}=-\frac{(36\pi)^{1/3}\sigma(T^4-T_{\rm irr}^4)}{s(\Delta\tau+1/\Delta\tau)}. 
\end{equation}
In the Equation above, $\sigma$ is the Stefan-Boltzmann constant, $T$ the gas temperature, $T_{\rm irr}$ the irradiation temperature, $s=(m/\rho)^{1/3}$ and $\Delta\tau=s\,{}k\,{}\rho$, where $m$ and $\rho$ are the gas particle mass and density and $k$ is the local opacity. 
The dust-to-gas ratio has been fixed to a constant value for each different dust species. For $k$, the adopted Planck and Rosseland dust opacities are taken from \citet{Dalessio01}. 
The irradiation temperature, which represents the minimum temperature allowed by the dust that acts as a thermostat for the gas, is set to $T_{\rm irr}=10 \, \mathrm{K}$.

No stellar feedback was included in this set of simulations, and we simply decided to assume that our clusters are the result of instantaneous gas removal at 3 Myr after the beginning of the hydro-dynamical simulation to roughly simulate the effect of the first supernova explosions. Indeed, \citet{Dale15} have shown that the pre-supernova gas removal is expected to play a minor effect on the survival and dynamics of stellar clusters and we also checked that at 3 Myr the gas accounts for a small fraction of the mass where most of the stellar mass is residing. Furthermore, at 3 Myr all the clouds converted about 30-40\% of their gas mass into sink particles, in agreement with previous hydro-dynamical simulations showing that stellar feedback should lead to a maximum star formation efficiency of about this amount \citep[e.g.,][]{VazquezSemadeni10, Dale15, Gavagnin17, Li19}.
For more details on such choices, we refer the reader to \citet{2020MNRAS.496...49B}.
\begin{figure*}
	\includegraphics[width=\textwidth]{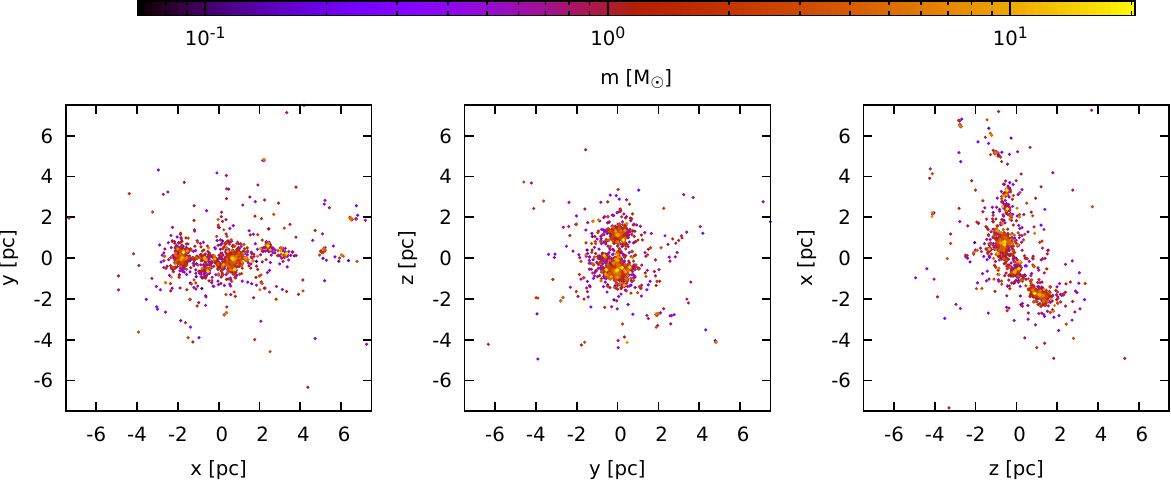}
    \caption{From left to right, projections in the $x-y$, $y-z$ and $z-x$ planes of the end state of the \texttt{m1e4} simulation. The colour map marks the mass of the individual stars in units of M$_{\odot}$.\label{projection}}
\end{figure*}

\subsection{Structural properties of the SPH simulations}

Independently of the specific initial value of $M_{\rm mc}$, our SPH simulations present a clumpy structure with $N_{\rm s}\approx 3\times 10^3$ stars\footnote{The approximately constant value of $N_{\rm s}$ follows from the fact that the star formation efficiency is roughly independent of $M_{\mathrm{mc}}$ (see Table 1 in \citealp{2020MNRAS.496...49B}). The star formation efficiency is indeed rather dictated by the physical processes involved in the simulations, which, in all cases, start from the same values of cloud temperature and density.}, organized in a maximum of $N_{\rm c}=9$ main sub-clumps for \texttt{m7e4} to a minimum of 2 for \texttt{m4e4}. Sub-clumps are identified heuristically as groups of neighbouring stars containing more than $0.05\,{} N_{\rm s}$, which self potential energy exceeds that of the rest of the system.
Figure \ref{projection} shows the $x-y$, $y-z$ and $z-x$ projections of the stars position on the three coordinate planes for the system  \texttt{m1e4}, with their masses $m$ shown in colour. We find a rather prominent primordial mass segregation, with heavier stars typically well within the central regions of the main clumps and lighter stars at larger distances from the geometric centres of such subsystems. 
All systems are above the virial condition, with $\alpha_{\rm vir}$ ranging from $1.19$ for the \texttt{m1e4} case, to $1.69$ for \texttt{m6e4}.

\begin{table}
\caption{Properties of the end states of the SPH simulations of \protect\cite{2020MNRAS.496...49B}.} 
\begin{tabular}{lllllll}
\hline
Name & $N_{\rm s}$ & $N_{\rm c}$ & $\alpha_{\rm vir}$ & $\gamma$ & $M_{\rm sink}$ $\left[{\rm M}_\odot\right]$ & $M_{\rm mc}$ $\left[{\rm M}_\odot\right]$ \\
\hline 
\texttt{m1e4} & 2523 & $6$ & $1.19$ & $2.30$ & $4.22\times 10^3$ & $10^4$ \\
\texttt{m2e4} & 2571 & $4$ & $1.32$ & $2.12$ & $6.69\times 10^3$ & $2\times 10^4$ \\
\texttt{m3e4} & 2825 & $5$ & $1.48$ & $2.20$ & $1.03\times 10^4$ & $3\times 10^4$ \\
\texttt{m4e4} & 2868 & $2$ & $1.47$ & $2.17$ & $1.44\times 10^4$ & $4\times 10^4$ \\
\texttt{m5e4} & 2231 & $4$ & $1.47$ & $1.80$ & $1.41\times 10^4$ & $5\times 10^4$ \\
\texttt{m6e4} & 3054 & $5$ & $1.69$ & $2.15$ & $2.04\times 10^4$ & $6\times 10^4$ \\
\texttt{m7e4} & 4214 & $9$ & $1.50$ & $2.20$ & $3.15\times 10^4$ & $7\times 10^4$ \\
\texttt{m8e4} & 2945 & $6$ & $1.60$ & $1.86$ & $2.83\times 10^4$ & $8\times 10^4$ \\
\texttt{m9e4} & 3161 & $4$ & $1.52$ & $1.90$ & $3.05\times 10^4$ & $9\times 10^4$ \\
\texttt{m1e5} & 3944 & $6$ & $1.46$ & $2.20$ & $3.80\times 10^4$ & $10^5$\\
\hline
\end{tabular}
\label{table:alpha}
\flushleft
\footnotesize{After the name of each simulation (Col. 1), we report the number of stars generated (Col. 2), the number of macroscopic subclumps (Col. 3), the virial ratio ($\alpha_{\rm vir}\equiv 2K/|W|$, Col. 4), the $\gamma$ coefficient of the mass-spectrum fitting function of (Eq.~\ref{tinypebbleshugeboulders}, Col. 5), the total mass of the stars (Col. 6), and the mass of the parent molecular cloud (Col. 7).}
\end{table}

\begin{figure*}
	\includegraphics[width=\textwidth]{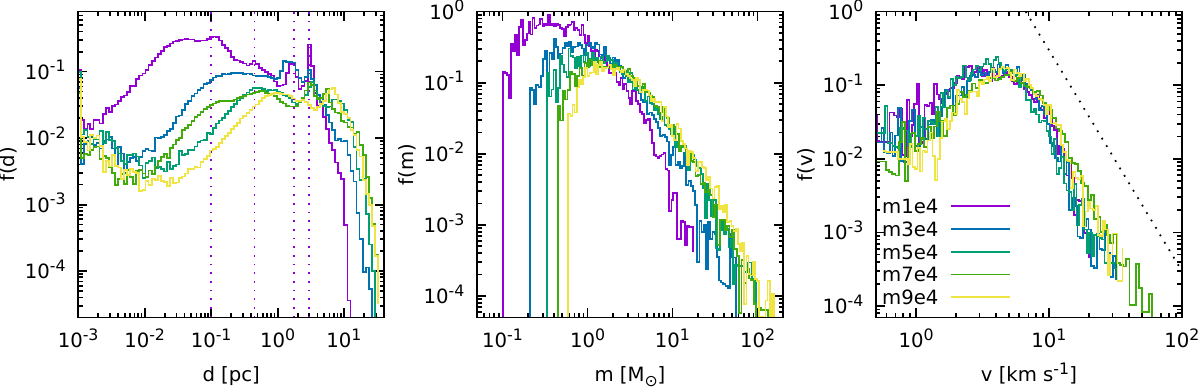}
    \caption{Distributions of inter-particle distances $f(d)$ (left-hand panel), mass spectra $f(m)$ (middle panel), and velocity distribution $f(v)$ (right-hand panel) for the sink particles taken from the simulations  \texttt{m$1$e$4$},  \texttt{m$3$e$4$}, \texttt{m$5$e$4$}, \texttt{m$7$e$4$}, \texttt{m$9$e$4$}. The vertical dotted lines in the left-hand panel mark the position of the main peaks of $f(d)$, corresponding to the distances between the main sub-clusters, for the \texttt{m1e4} case. The thin dotted line in the right-hand panel marks the $v^{-3}$ power-law trend of the velocity distributions. \label{distances}}
\end{figure*}

In order to quantify the properties of the end states of the SPH simulations, we have evaluated their distributions of inter-particle distances $f(d)$, mass spectra $f(m)$, and velocity distributions $f(v)$. Figure \ref{distances} shows these distributions for the sink particles of the simulations \texttt{m1e4}, \texttt{m3e4}, \texttt{m5e4}, \texttt{m7e4} and \texttt{m9e4}. 
The distribution of inter-particle distances 
shows a quite complex structure with several slope changes. The clumpy structure of the particles' spatial distribution gives rise to several peaks in $f(d)$, corresponding to the distances between the clumps themselves. 
For the specific case of \texttt{m1e4}, the peaks are located roughly at 0.1, 0.45, 1.75 and 3 pc (as highlighted by the vertical dotted lines), that can be identified as the distances between the approximate centres of the main clumps of the particles shown in Fig. \ref{projection}.

The mass spectra of sink particles approximately follow the same power-law structure between a low-mass and a high-mass cut-off. The differences in the lower mass limit are due to the different mass resolution of the hydro-dynamical simulations, which are initialized with different total masses but the same number of particles, as explained in Sect. \ref{sec_set_up}. At higher masses, where the physical processes involved in the simulation become the dominant factor in shaping the mass function, all the spectra recover the same slope.  We have fitted the numerically recovered mass spectra with the {\it bona-fide} function:
\begin{equation}\label{tinypebbleshugeboulders}
f(m)=\frac{C}{\left(m^2+m_*^2\right )^{\gamma/2}},
\end{equation}
where $C$ is a normalization constant, $m_*$ is a scale mass, that for the explored systems is always in the range between 0.8 and 4 M$_{\odot}$, while exponent $\gamma$ ranges from $\approx 1.8$ to $\approx 2.3$.

The velocity distributions $f(v)$ do not show a relevant dependence on the specific initial value of $M_{\rm mc}$, as shown in the right hand panel of Fig. \ref{distances}. Qualitatively, the velocity distribution is well described by a Maxwell-Boltzmann distribution from $v=0$ to 5~km s$^{-1}$ (value corresponding to the peak of $f(v)$) and then shows a $v^{-3}$ power-law trend. The properties of the SPH simulations are summarized in Table~\ref{table:alpha}.

\section{Methods}
\label{meto}

In the following, we describe our new procedure to build a generative model of star cluster initial conditions. 
In principle, a generative model's goal is to learn a representation of an intractable distribution given an usually finite number of samples. The generator typically maps from a latent domain on which a simple distribution is defined, such as a multivariate Gaussian on $R^n$, to the complex data domain \citep[e.g.][]{2021arXiv210305180R}.
Recently, most of the interest in generative models is driven by deep learning approaches, such as generative adversarial networks \citep[][]{2014arXiv1406.2661G}. However, in principle, much simpler models such as hidden Markov models \citep[][]{1165342, eddy2004hidden} or grammars \citep[e.g.,][]{CHOMSKY1959137, 10.1007/978-3-642-76626-8_35, beaumont2009grammatical} meet the definition of generative model in the broader sense defined above. The latter have proved useful in the description and generation of objects displaying fractal structure, as in the case of Lindenmayer systems applied to plant growth \citep[][]{LINDENMAYER1968280, LINDENMAYER1968300,  prusinkiewicz2013lindenmayer}.

Our generative approach focuses on reproducing the complex fractal structure of embedded star clusters from hydro-dynamical simulations (see, e.g., Fig 4 in \citealp{2020MNRAS.496...49B}) by capturing the relations between sub-clusters at different scales through a hierarchical clustering algorithm. This will eventually allow us to generate new realizations by modifying their macro structure, i.e. the relations between large sub-clusters. The parameters that characterize the relevant properties of these clumps and their relations can be treated as the latent domain of our generative model. 

We proceed in two steps. First, 
we use a hierarchical clustering algorithm to identify clumps of stars at different scales in the phase space of the original hydro-dynamical simulation output. The clumps are organized by the algorithm into a  hierarchical tree $\mathcal{T}$, where the root node contains the whole set of stars and each subsequent node represents a two-way split with each branch being a clump of stars, down to the leaf nodes representing individual stars. 
For each node $\mathcal{T}_i$, we describe the relevant physical properties of the cluster in terms of the distance vector between the centres of mass of the clumps $\mathbf{l}_i$, their relative velocity vector $\mathbf{u}_i$, and the mass ratio between the two clumps. To describe how the mass is split at each node we refer to {\bf $q_i$}, defined as the ratio between the lightest of the two resulting groups and the total mass of the node. With this definition, mass ratios fall between $0$ (maximally unequal split) and $0.5$ (equal-mass split). The description of the star clusters in terms of the hierarchical clustering algorithm is given in Sect. \ref{sec_application_to_sc},  but its goal in short is to capture structure as a function of scale, similarly to what was done in, e.g., \cite{elmegreen06} by applying smoothing kernels of different sizes.

Second, we generate a new realization of particle positions and velocities by placing clumps of stars (and sub-clumps down to the individual stars) in phase space. 
To build a new realization of total mass $M$ (details in Sect. \ref{sec_generative_model}), we start with one particle at rest in the origin of our coordinate system, initially containing the total mass of the cluster $M$. Then, we iteratively split it into new particles and place them, at each step $i$, at a distance  $\mathbf{l}_i$ from each other, moving with relative velocity $\mathbf{u}_i$.  
The relevant variables $\mathbf{l}_i$, $\mathbf{u}_i$, and the relevant mass ratio $q_{i}$ are taken from the tree $\mathcal{T}$  except for the first step(s), which are drawn from a tree $\mathcal{T}^\prime$ built on a different simulation. While this does not guarantee that the outcome will be described by a tree with statistical properties that match those of $\mathcal{T}$, it is at least heuristically convincing in the case of very hierarchical distributions. Moreover, we will check ex post that the realizations generated in this way have a set of desirable properties with respect to the original cluster. The details about the generative procedure are given in Sect. \ref{sec_generative_model}.

\subsection{Hierarchical clustering} \label{hierarchical_clust}


Hierarchical clustering algorithms arrange data into a tree-like structure representing nested groups, capturing clustering structure at different scales. In particular, we use an agglomerative clustering algorithm \citep[see the chapter on {\sc agnes} in][]{1990fgda.book.....K}. This means that the tree-like hierarchy of clusters is built from the bottom up: the algorithm starts from individual points, and merges the most similar ones into clusters until some stopping criterion is satisfied (e.g., until only a specified number of clusters are left). This way of proceeding can be thought as drawing a tree with a branch for every pair of clusters that merge\footnote{
The procedure of drawing a hierarchy of merging sub-structures may recall the merger tree history, which is used in cosmology to track the assembly of sub-structures across time (see, e.g., \citealp{rodriguezgomez15}). 
Agglomerative clustering algorithms, however, do not imply any evolution in time, but use the tree-like structure to identify groups of instances at different scales.
}. A dendrogram can be used to display the resulting tree structure, with leaf nodes corresponding to individual points and the root corresponding to the whole data set. We refer the interested reader to \cite{2019arXiv190604983P} for an illustration of this and other clustering algorithms in an astronomical context. 
Here, we selected this algorithm because it is well suited for studying the complex structure of the hydro-dynamical simulations described in Section~\ref{simu}, since it is informative on very different scales and it can capture clusters (and sub-clusters) of various sizes. We use the implementation offered by the {\sc scikit-learn} library \citep[][]{scikit-learn}\footnote{The details about the implementation of the algorithm can be found at \href{https://scikit-learn.org/stable/modules/generated/sklearn.cluster.AgglomerativeClustering.html}{this link}.}.

\begin{figure*}
	\includegraphics[width=\textwidth]{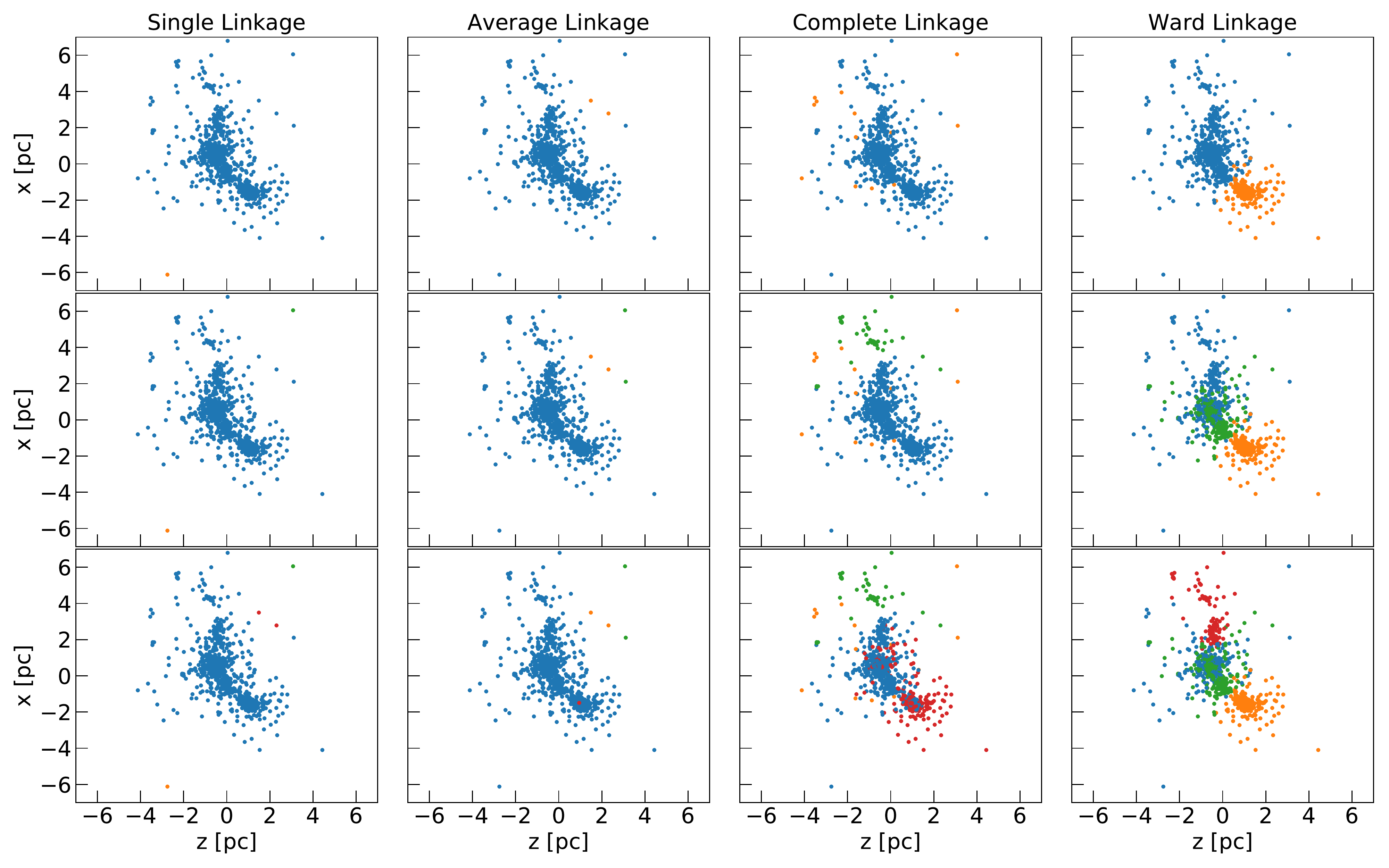}
    \caption{ First nodes from the trunk in the hierarchical tree for the \protect\texttt{m1e4} simulation, obtained by considering different linkages: single (first column), average (second column), complete (third column), and Ward (last column) linkage. The panels in the first row show the first node of the tree, that splits the sink particles of the simulation into two groups (blue and orange). In the second node, the blue group is split further into two subgroups (blue and green, panels in the second row). The third node splits the blue group into the blue and the red groups (panels in the lower row).} \label{fig_linkage}
\end{figure*}

\subsubsection{Linkage}

Moving towards the root of the tree, an agglomerative clustering algorithm merges at each node either two groups with each other or a lone point into a group. This process is based on a notion of (dis)similarity between groups which may be defined in multiple ways, or linkages. 
We considered four different linkages and evaluated their performance in clustering the sink particle spatial distribution.
\begin{itemize}
    \item The \emph{single linkage} merges the two clusters that have the minimum distance between any points in the two groups:
    \begin{equation}
    \Delta_{AB} \coloneqq \min{(l_{{i \in A},\,{}{j \in B}})},
    \end{equation}
    where $i$ and $j$ represent sink particles belonging to group $A$ and $B$, respectively, and  $l_{i,j}$ is the distance between two such particles. 
    \item The \emph{average linkage} merges the two clusters that have the smallest average distance between all their points:
    \begin{equation}
    \Delta_{AB} \coloneqq {\rm mean}{(l_{{i \in A},\,{}{j \in B}})}. 
    \end{equation}
     
    \item The \emph{complete linkage} (also known as maximum linkage) merges the two clusters that have the smallest maximum distance between their points:
    \begin{equation}
    \Delta_{AB} \coloneqq \max{(l_{{i \in A},\,{}{j \in B}})}.  
    \end{equation}
    
    \item \emph{Ward's linkage} merges two clusters such that the variance within all clusters increases the least. This often leads to clusters that are relatively equally sized. Ward's linkage is defined as follows:
\begin{equation}\label{eq:ward2}
    \Delta^2_{AB} = \sum_{i \in {A \cup B}} l^2_{i, c_{A \cup B}} - \left(\sum_{i \in A} l^2_{i, c_A}  + \sum_{i \in B} l^2_{i, c_B} \right),
\end{equation}
where the index $i$ denotes the generic $i-$th particle and $c_A$, $c_B$, and $c_{A \cup B}$ denote the centroids of sets $A$, $B$, and ${A \cup B}$ respectively. Equation~\ref{eq:ward2}
corresponds to the increase in variance with respect to the relevant centroids as groups $A$ and $B$ are merged. Merging groups decreases the number of centroids by one, so variance is bound to increase, but using Ward's linkage results in cluster mergers that minimize its increase at each step. 

\end{itemize}

Figure \ref{fig_linkage} shows how the choice of the linkage affects the structure of the first three nodes of the tree of \texttt{m1e4}. The single linkage approach leads to a single, big sub-clump, separated from a few isolated stars. In fact, following this prescription, two blobs that just touch in one point are considered similar and get merged into one pretty quickly, even if their centres-of-mass are far from each other. In contrast, single isolated stars are merged only in the final branches. The average and complete linkage perform poorly as well, likely because their merging criterion is too simple to fit the complex structure of the hydro-dynamical clusters. Finally, Ward's linkage performs well in describing the large scale structure of the cluster, as it correctly identifies the main clumps and is thus informative about the structure of the cluster. For this reason, hereafter we will consider only  Ward's linkage.


\begin{figure}
	\includegraphics[width=\columnwidth]{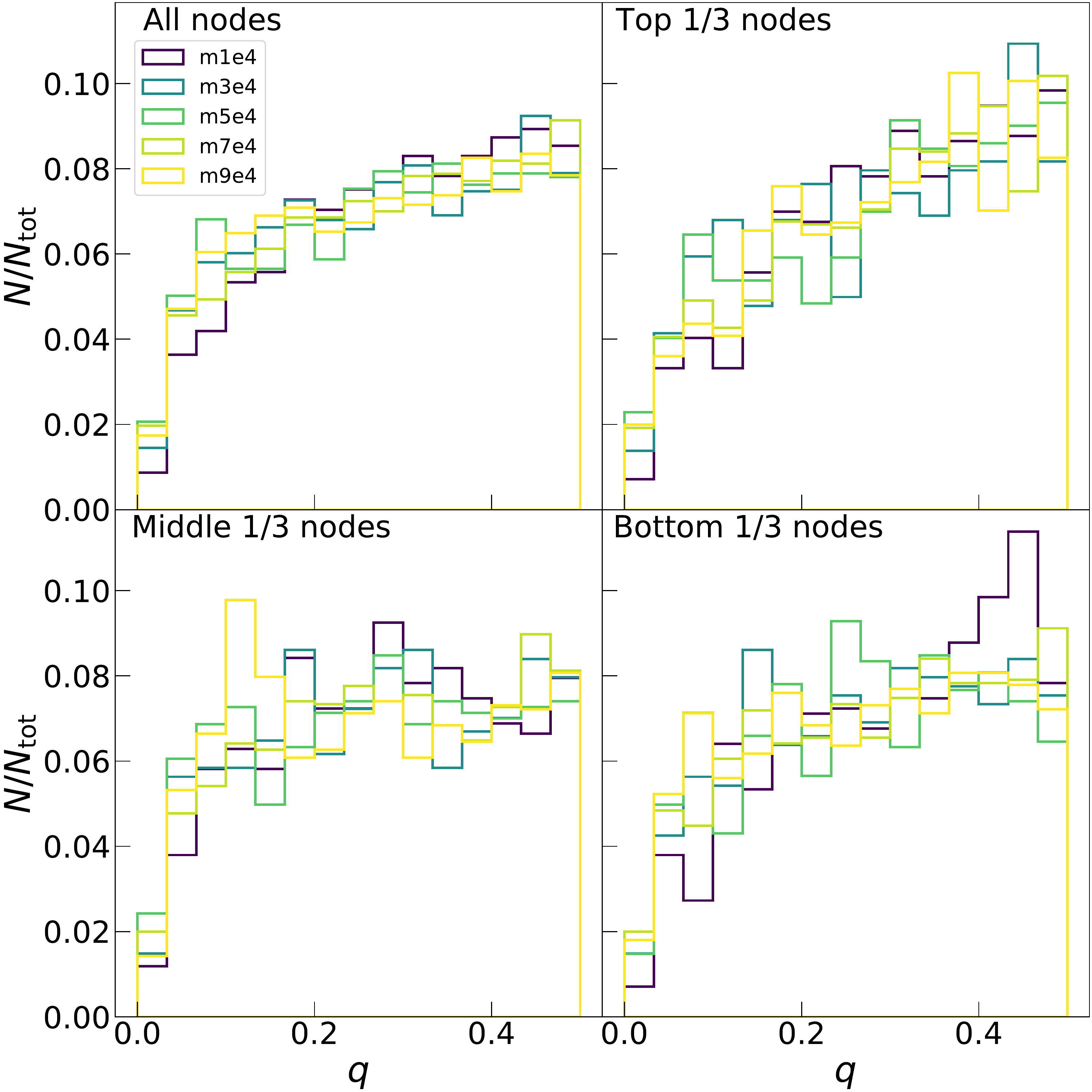}
    \caption{Distribution of the mass of the lightest of the two resulting groups at any given split, in units of the parent group. The top left panel shows the distribution calculated for all nodes in the learned tree. The top right panel shows the distribution for the top $1/3$ of the nodes from the root (big clumps), the bottom left for the middle $1/3$ of the nodes (intermediate-size clumps), and the bottom right for the lower $1/3$ of the nodes (small clumps to individual stars). \label{mass_splitting_histograms}}
\end{figure}

\subsection{Application of hierarchical clustering to stellar clusters} \label{sec_application_to_sc}

We applied agglomerative clustering to the stellar clusters from hydro-dynamical simulations introduced in Section~\ref{simu}. The trees are built by relying on euclidean distance between sink particles in the phase space as a measure of dissimilarity, so that particles sharing both similar positions and similar velocities tend to be grouped together. Before applying the algorithm, we scaled the positions and the velocities by their standard deviations. This step or some such is necessary so that the result of our clustering does not depend on the arbitrary choice of the unit of measurement of time.

The right column of Fig.~\ref{fig_linkage} shows the groups of sink particles corresponding to the first two nodes of the learned tree (starting from the root).
The first node splits the sinks into two big chunks, and the second node splits off a smaller clump from one of these\footnote{
Even as we describe the tree from the root up (writing occasionally in terms of splits/splitting) agglomerative methods build the tree from the leaves, i.e. the individual sink particles.}. Our choice of using Ward linkage results in the splitting off of the most massive sub-clumps in the first branches of the tree, leading to an overall balanced tree. The first splitting thus gives information about the distribution of the sub-clumps at large scales and, moving towards the leaves of the trees, sub-clusters are split in smaller and smaller sub-clumps, as desired for our task.

\begin{figure}
	\includegraphics[width=\columnwidth]{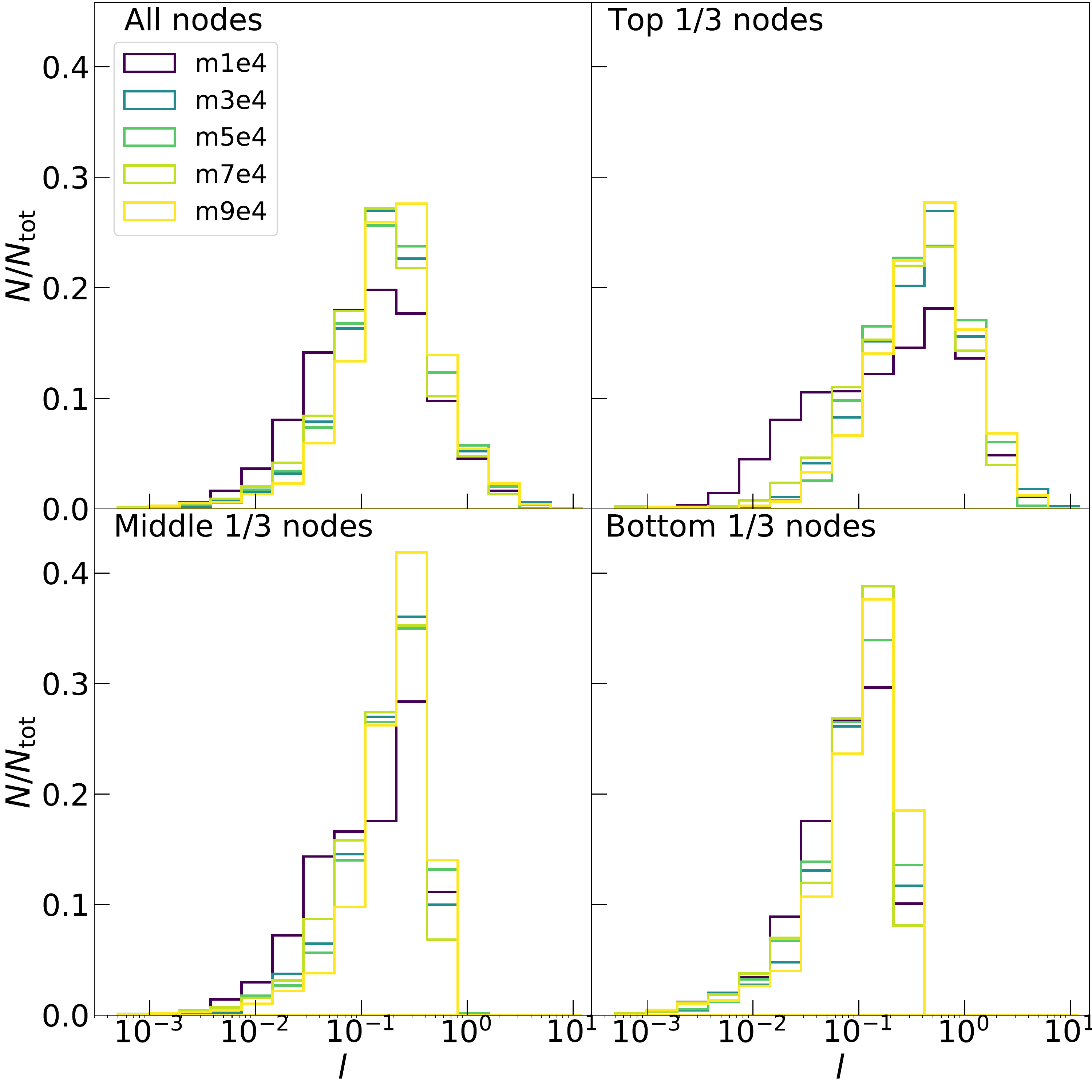}
    \caption{Same as Figure~\ref{mass_splitting_histograms} but for the distribution of the distances (scaled by their variance) between the centres-of-mass of two resulting groups at any given split 
    of the agglomerative clustering hierarchical tree. 
    } \label{cm_histograms}
\end{figure}

Figure~\ref{mass_splitting_histograms} shows the mass ratios between sub-clumps branching off at different depths within the tree. 
The distribution of mass ratios is not particularly affected by the tree depth. This is expected if the structure of the sink particle distribution is scale invariant, as moving down the tree (towards the leaves) probes smaller scales by construction.
Additionally, Fig.~\ref{mass_splitting_histograms} shows that the distribution is similar across different simulations, spanning a range of total mass of an order of magnitude. 
To assess if the mass ratios can be considered as drawn from the same distribution (after properly rescaling the mass), we performed pairwise Kolmogorov--Smirnov tests. Despite multiple testing we never obtain a p-value below $10^{-2}$, so we have no reason to suspect that the distributions are different. 
Also, we performed the same test on the sub-distributions shown in Fig.~\ref{mass_splitting_histograms} separately. Our test always obtains p-values above $10^{-1}$, with the only exception for the comparison between the middle nodes of \texttt{m1e4} and \texttt{m9e4}, where p-value~$=10^{-1.2}$. 
This result suggests that,  despite some statistical fluctuations, the splitting in mass is performed in the same way at different scales for all the simulations.  

\begin{figure}
	\includegraphics[width=\columnwidth]{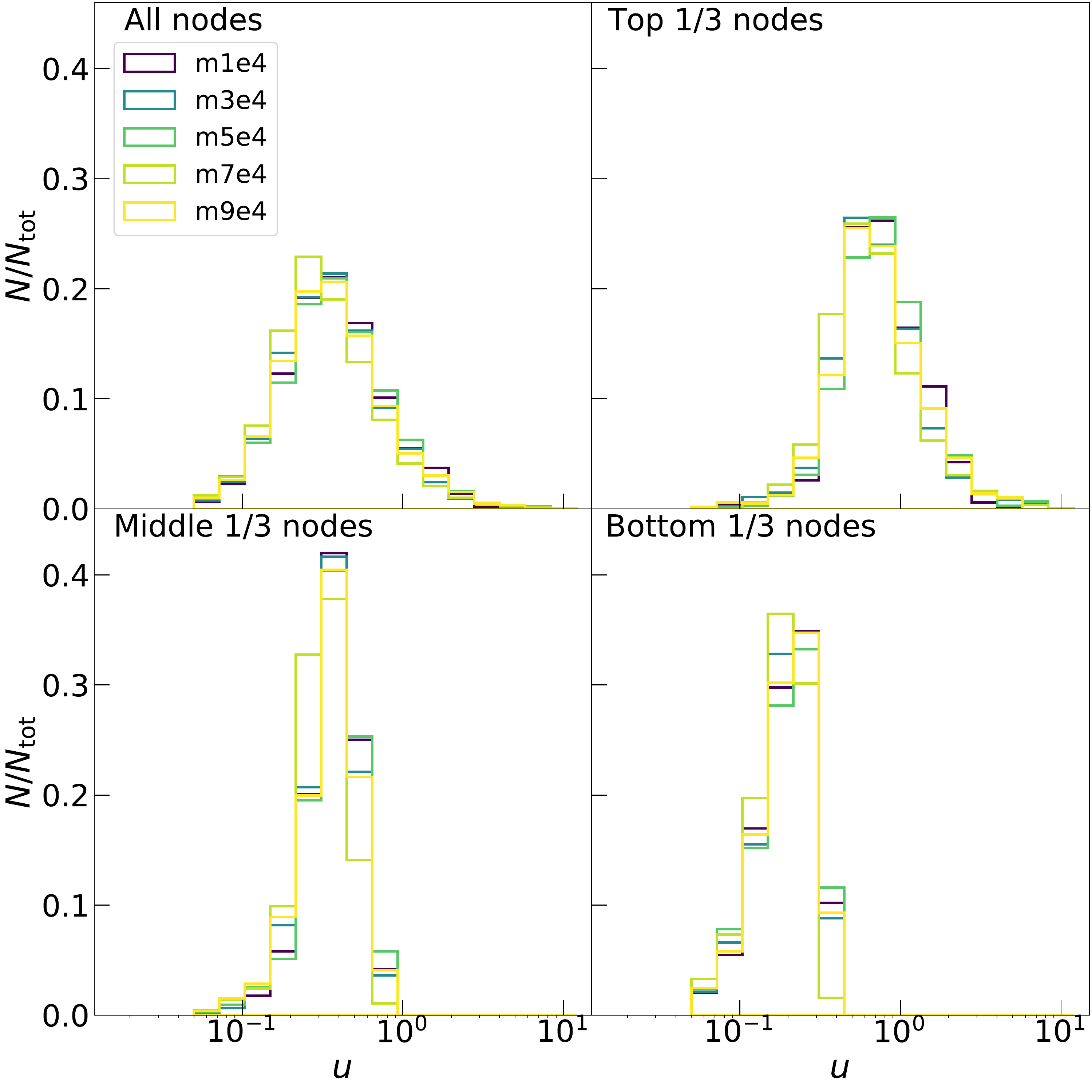}
    \caption{Same as Figure~\ref{mass_splitting_histograms} but for the distribution of the relative velocities (scaled by their variance) between the centres-of-mass of two resulting groups at any given split 
    of the agglomerative clustering hierarchical tree. 
    }\label{vcm_histograms}
\end{figure}

\begin{figure}
	\includegraphics[width=\columnwidth]{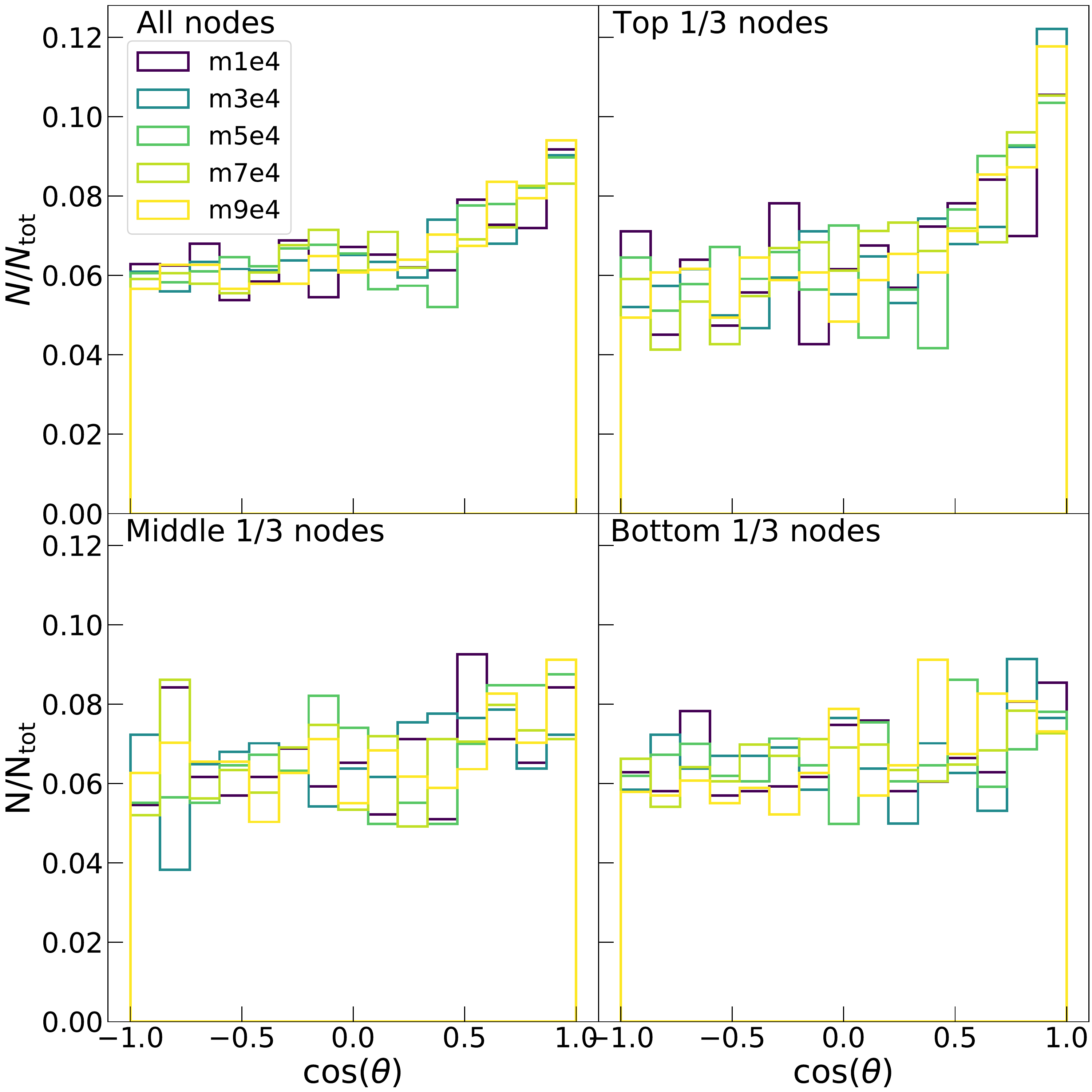}
    \caption{Same as Figure~\ref{mass_splitting_histograms} but for the distribution of the cosine of the angle between the relative velocity and the distance of the centres-of-mass of two resulting groups at any given split 
    of the agglomerative clustering hierarchical tree.  
    \label{theta_histograms}}
\end{figure}

Similar information on the scaling behavior of our simulations can be extracted from Figures~\ref{cm_histograms} and \ref{vcm_histograms}, where we show the distribution of the distances between the clumps ($l=|\mathbf{l}|$) and that of their relative velocities ($u=|\mathbf{u}|$). In particular, the positions of the maxima of the distributions shift towards lower values by moving from the top to the bottom nodes, confirming that the tree is considering smaller and smaller scales. Also in this case, all the simulations show very similar distributions at each level for both the distances and the relative velocities. 
The distribution of the angles between the relative velocity and the distance, $\theta = \arccos{( \mathbf{l} \cdot \mathbf{u} \, (l \, u )^{-1})}$, is shown in Figure \ref{theta_histograms}. 
This distribution appears flat except for a rise at $\cos{\theta} \approx 1$ which corresponds to relative velocity parallel to the separation vector between clumps, which is expected in a super-virial cluster undergoing overall expansion.

\begin{figure}
	\includegraphics[width=\hsize,keepaspectratio]{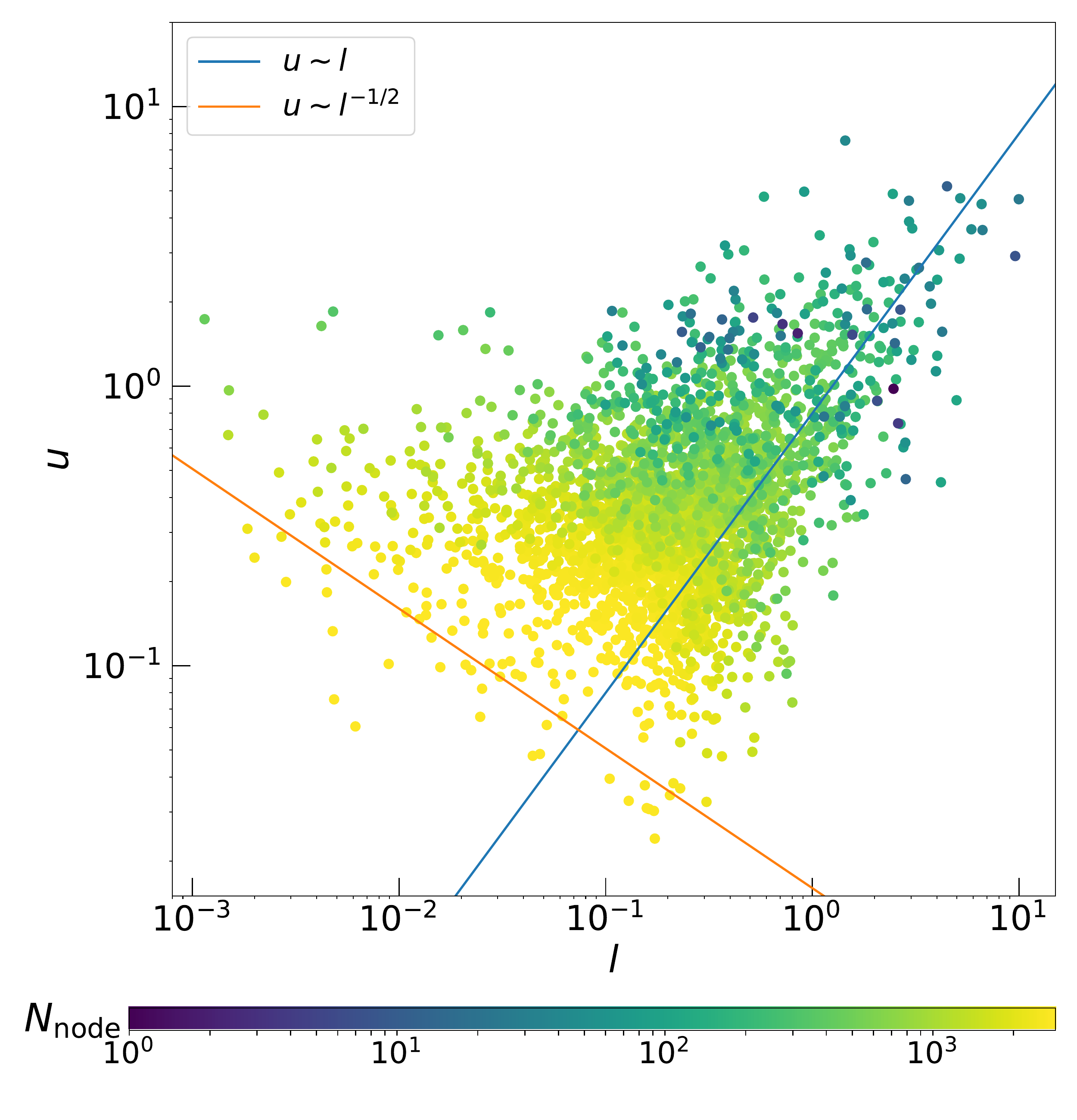} 
    \caption{Scatter plot of the relative velocity between the centres-of-mass of two different sub-clumps corresponding to a given node in the 
    agglomerative clustering tree as a function of their  distance. The colour gradient maps the depth of the node (from the root, in blue, to the leaves, in yellow) within the hierarchical tree, $N_{\rm node}$. The superimposed lines represent two limit slopes corresponding to rigid rotation (blue) and Keplerian motion (orange). 
    \label{figsscatter}}
\end{figure}

Relevant physical information can be drawn by considering the relation between quantities of the same node in the 
agglomerative clustering tree. Figure \ref{figsscatter}  shows the relation between the distance of the sub-clumps and their relative velocity, for each node. The main sub-clumps, that correspond to the nodes closest to the root, show a direct proportionality between these two quantities, possibly due to rigid rotation. In contrast, on the smallest scales, the single particle relative velocity shows a tendency to decline with the square root of their distance, as would happen for two clumps (or even two individual stars) orbiting one another under the influence of each other's monopole potential. Interestingly, all relative motions between clumps take place between the rigid and Keplerian extremes.



\subsection{Generating new realizations} \label{sec_generative_model}
\begin{figure*}
	\includegraphics[width=\textwidth]{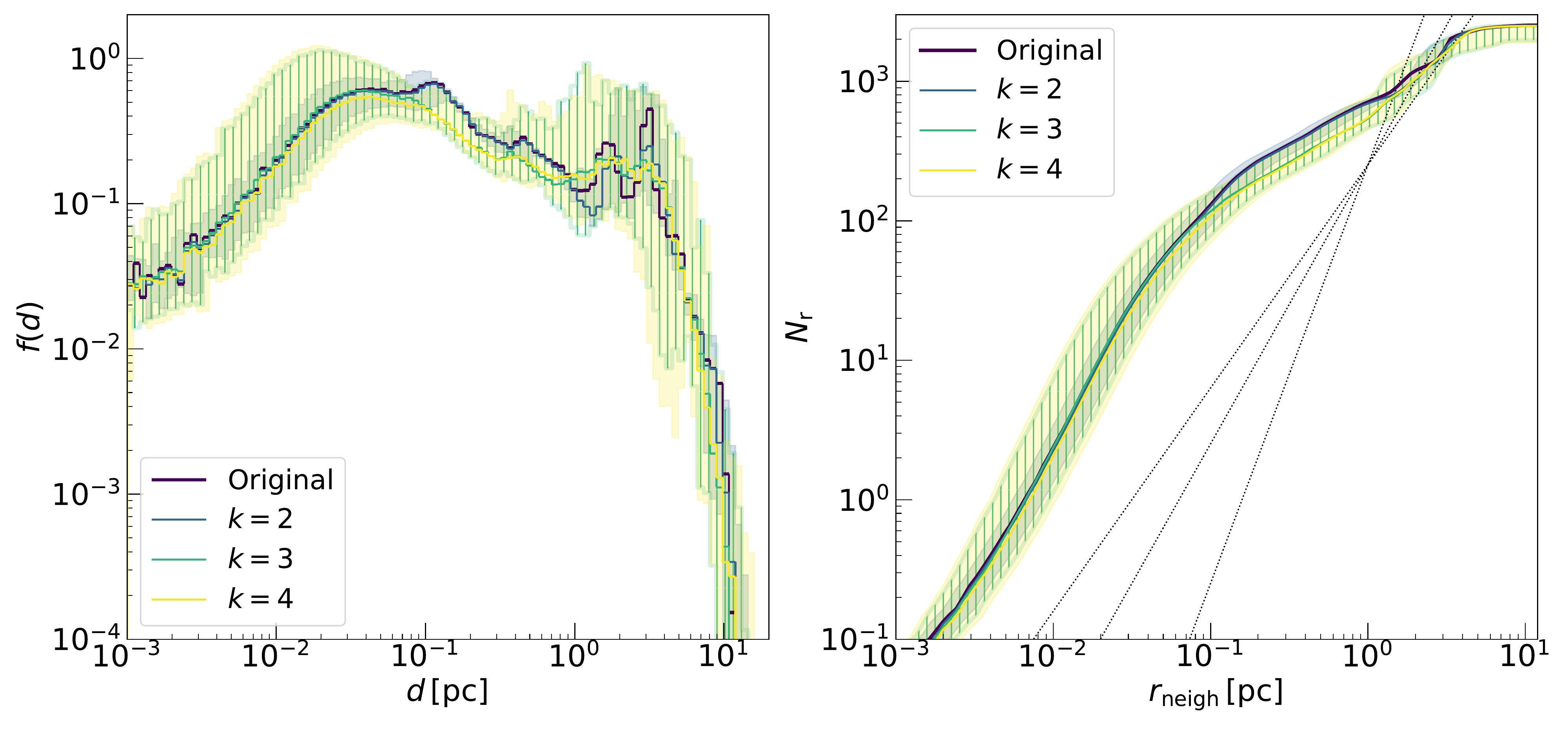}
    \caption{Left-hand panel: distribution of inter-particle distances $f(d)$ for the sink particles taken from the \texttt{m$1$e$4$} simulation (thick purple line) and three distributions of new generations obtained by replacing the first 1 (blue), 2 (green, hatched area) and 3 (yellow) nodes, corresponding to $k=2$, $3$, and $4$ in the notation used above. The shaded area encloses the distribution of the new generations, and the solid line is the median of the distribution.
    Right-hand panel: average number of neighbours $N_{\rm r}$ around a star, within a sphere with radius $r_{\rm{neigh}}$, for different values of $r_{\rm{neigh}}$. Lines and colors are the same as in the left-hand panel. The black dotted lines represent the trend expected for distributions with a uniform fractal dimension, for $\beta=1.6$, $2$, and $3$ 
    .} \label{fd}
\end{figure*}

As explained in Section~\ref{sec_application_to_sc}, the application of the agglomerative clustering algorithm to stellar clusters allows us to inform a tree $\mathcal{T}$ encoding their hierarchical structure. Each node of the three $\mathcal{T}_i$ is associated to the relevant properties $\mathbf{l}_i$, $\mathbf{u}_i$, 
and $q_{i}$, which quantify the relations between the sub-clumps corresponding to the branches departing from the node. Thus the tree essentially encodes instructions to generate a new star cluster, as it can be traversed from the top, iteratively splitting an intial particle until the leaf level is reached, where individual stars have been produced.
In our case, the goal is to change the cluster at the global structure level -nearest to the trunk of the tree-, thus creating different sub-clumps configurations while preserving the small scale properties of the sub-clumps (such as their fractal structure).
We thus take the quantities $\mathbf{l}_i$, $\mathbf{u}_i$, 
and $q_i$ associated to the nodes $\mathcal{T}_i$ for $i < k$ and 
replace them with the quantities  $\mathbf{l}^\prime_i$, $\mathbf{u}^\prime_i$, 
and $q^\prime_i$ associated to the nodes $\mathcal{T^\prime}_i$ of another tree $\mathcal{T^\prime}$, learned from a different set of sink particles. 
This grafting procedure represents a way to combine the large scale properties of one simulation with the small scale properties of another. 
For the results presented in Sect. \ref{resu}, these nodes are sampled randomly from other simulations. 

The generation procedure is implemented as follows. First, we consider a particle with a mass $M_1$ equal to the total mass of the cluster considered, placed at the centre of mass of the cluster. 
The particle is first split into two particles of masses ${M_1}_1$ and ${M_1}_2$ such that ${M_1}_1 + {M_1}_2 = M_1$ and $\min({M_1}_1, {M_1}_2)/M_1 = q^\prime_1$. The positions and velocities of the new particles are assigned such that their centre of mass is at rest in the origin of the system, their distance vector is $\mathbf{l}^\prime_i$, and their relative velocity $\mathbf{u}_1^\prime$
. This splitting procedure is then repeated until a cluster with the same number of particles as the reference one is obtained. At each step, the particle-to-split is chosen by considering the same order of splitting as the original reference tree.
This procedure may at times result in very low mass particles. We remove these planet-sized objects with a cutoff at the minimum mass of the original stars on which $\mathcal{T}$ was learned. 

\subsubsection{Grafting depth}
In the procedure described above, the choice of the grafting depth $k$ determines how different the new realizations are from the original system. A 
low value of $k$ produces generations that are very similar to the original one at all scales. In contrast, when $k$ is 
high, also the small scales are modified substantially. In our case, we want to generate new clusters that are similar to the original one but, at the same time,  cannot be considered as its copies.
We evaluated how the choice of the grafting depth affects the spatial structure of the new generations. In particular, we considered the distributions of distances and the fractal dimensions obtained by generating sets of one hundred new realizations for \texttt{m1e4}, at different values of $k$. The left panel of Fig.~\ref{fd} shows the general shape of the distributions of inter-particle distances. Predictably, the realizations obtained by modifying just one node match the original distribution better than those that change two or three nodes, and present the smallest spread. The  peaks correspond to sub-clumps of sinks, that are formed in different numbers and sizes in each realization. 
At small distances, the new realizations recover the the general trend of the original distribution, as meant for our method. For the case with $k=1$, this happens at about 1 pc, meaning that only the very large scales (the distance between the main sub-clumps) are modified. By increasing $k$, also the smaller scales are altered, and the original shape is recovered later. This suggests that very few changes are sufficient to produce generations that can be defined as different from the original cluster.
The distributions for $k\leq3$ are consistent with the original simulation throughout the range of distances. 

\begin{figure}
	\includegraphics[width=\hsize]{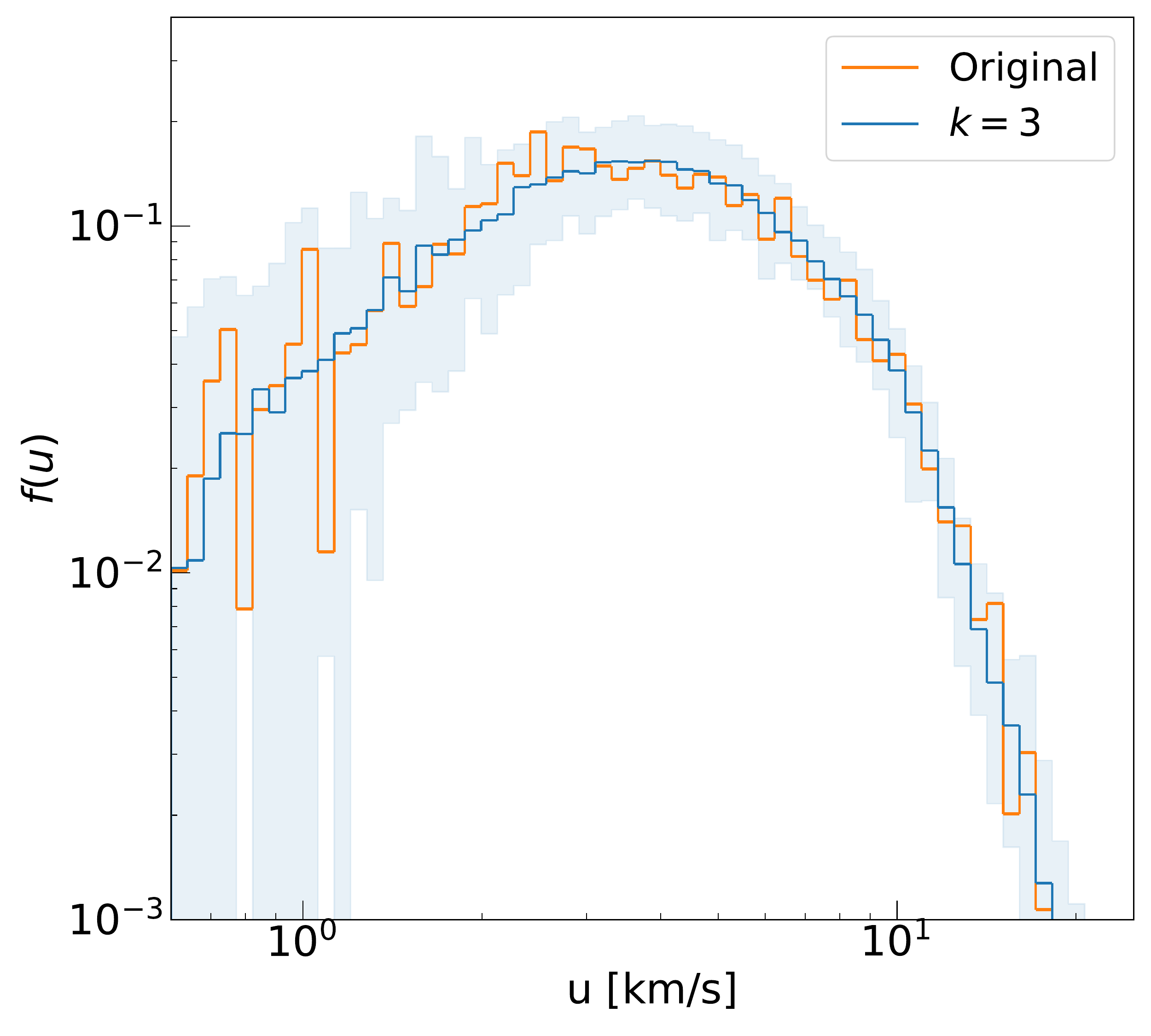}
    \caption{Distribution of velocities $f(u)$ for the sink particles taken from the \texttt{m$1$e$4$} simulation (orange line) and for a distribution of new generations obtained by considering $k=3$ (blue). The shaded area encloses the distribution of the new generations, and the solid line is the median of the distribution. 
    }\label{fig_velocity_generations}
\end{figure}

In the right-hand panel of Fig.~\ref{fd}, we have computed the fractal dimension by means of the average number of neighbours of the stars within a given distance, following \cite{2020MNRAS.496...49B}. The distribution of neighbours of \texttt{m1e4} is not described by a single power law of non integer index $\beta$, as one would expect in a simple fractal structure, but presents two slope changes at around $\approx 10^{-1}$ and $\approx 2$ pc (see also  \citealt{2020MNRAS.496...49B}). To guide the eye, the three dotted lines mark the theoretical distance distributions in the case of a pure fractal distribution with $N_{\rm r}\propto r_{\rm neigh}^{\beta}$, 
for $\beta=1.6$, $2$, and $3$.  
The generated distributions match the general trend and the changes in the slope of the original simulation very well, showing that our method captured the underlying structure of the particle distribution in the 3D space at all scales. 
Like for the inter-particle distance distribution, the choice of $k=2$ produces only minimal differences from the original \texttt{m1e4} profile. In the following, we will focus on generations with $k=3$, which allows to produce a distribution of clusters that are distinguishable but still consistent with the original one at all scales.

Figure~\ref{fig_velocity_generations} shows the distribution of velocities for the new generations obtained by setting $k=3$, as compared to the original sink particle trend. The median of the new generations matches the original distribution at all velocities, both on the low-velocity tail, where the Maxwell-Boltzmann trend seems to be preserved, and on the sharper power-law trend at high velocities.
At very low values ($u<1 \, \mathrm{km/s}$), the very low number of stars causes large fluctuations in the distribution of new generations, but their median trend is still well consistent with the original one.

\begin{figure*}
	\includegraphics[width=0.9\textwidth]{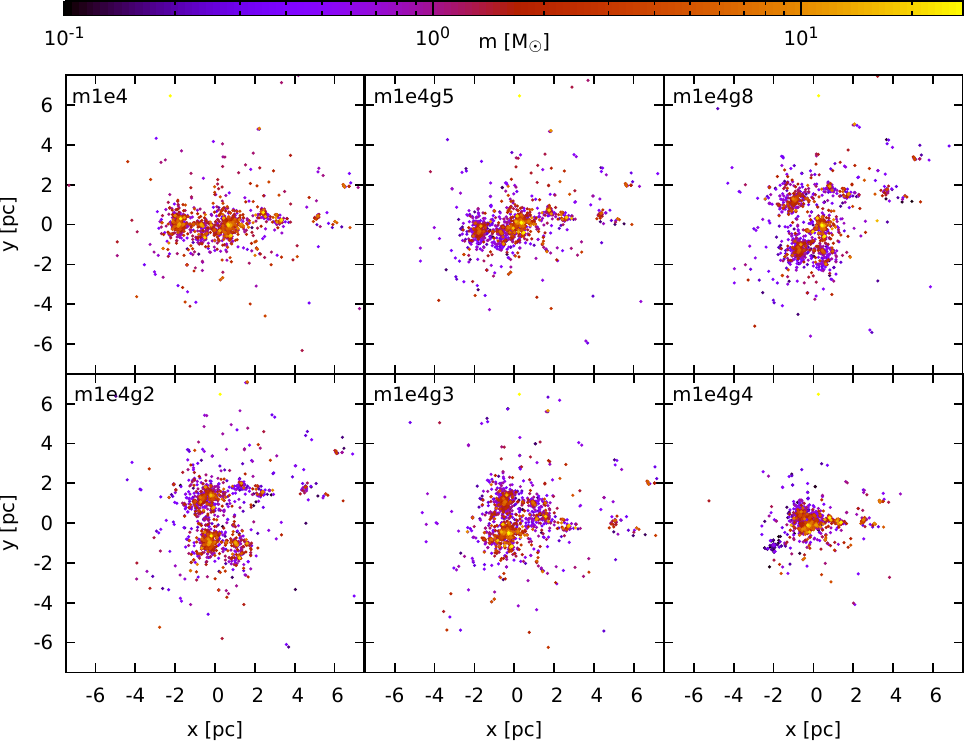}
    \caption{$x-y$ projection of the \texttt{m1e4} system (top left panel, see also Fig. \ref{projection}) and five different new
    generations. The colour code marks the different masses of the sink particles and their new generations.\label{regeneration}}
\end{figure*}

\begin{figure}
	\includegraphics[width=0.9\columnwidth]{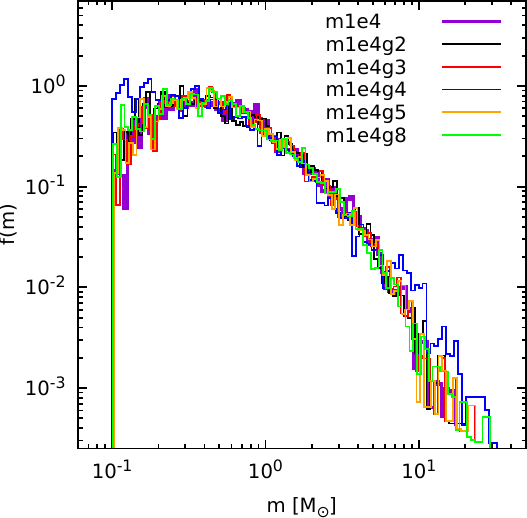}
    \caption{Mass spectrum of the sink particles of the simulation \texttt{m1e4} (thick purple line) and of five different generated systems (thin lines). \label{new_mass_histograms}}
\end{figure}
\section{Results}
\label{resu}

\subsection{Properties of the newly generated systems}

In this Section, we discuss the properties of the systems generated using our procedure starting from the simulation \texttt{m1e4}, which presents the highest resolution.
\indent Figure \ref{regeneration} shows the spatial distributions of five new generations obtained with the method described in Section~\ref{sec_generative_model}, compared to the original one. 
The new generations show a strong sub-structured configuration, with a different number of clumps, depending on the single realization, which has drawn branches from different simulations. Also, a strong degree of mass segregation is still present in the single sub-clumps, as highlighted by the colour coding. This primordial mass segregation in the individual realizations qualitatively matches the one present in the original cluster.

In Fig.~\ref{new_mass_histograms}, we compare the mass distribution of \texttt{m1e4} to those of the new generations. In this case, our method leaves the slope of the mass function largely unaltered for most of the mass spectrum. At the boundaries of the mass spectrum, some discrepancies are present. This is due to the fact that the change in the first nodes may split up a relatively small particle more times than in the original cluster, and leaves more massive particles less split. This explains the higher number of particles at the boundaries of the mass spectrum with respect to the original one. The sharper cut-off at $m\approx{10^{-1}}\,{}{\rm M}_{\odot}$ is due to the fact that all masses below this threshold are systematically removed.
In general, the fit with Eq.~\ref{tinypebbleshugeboulders} is rather good, yielding values of $\gamma$ around $2.3$, reminiscent of a \cite{1955ApJ...121..161S} slope. 
\begin{figure*}
	\includegraphics[width=\textwidth]{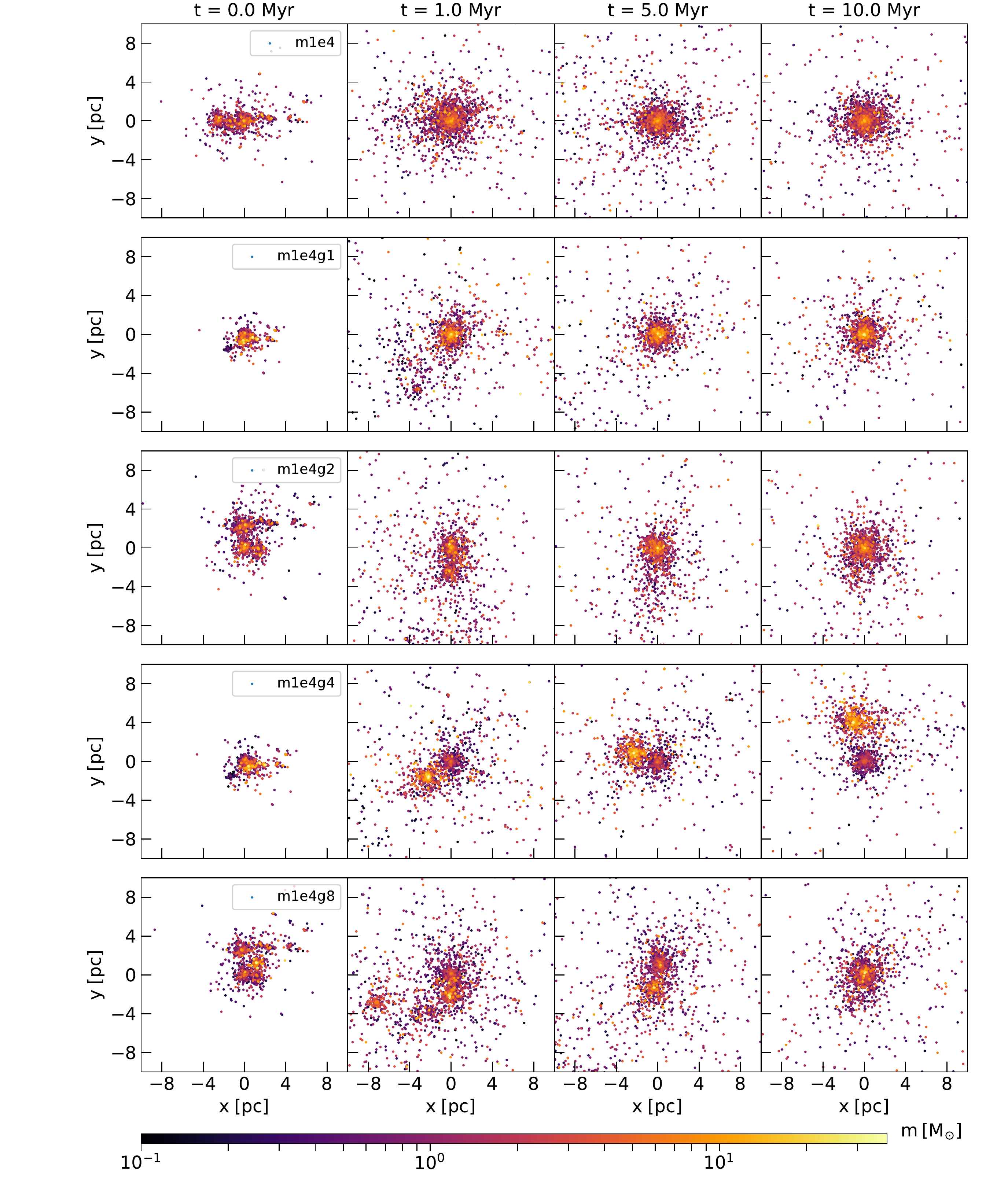}
    \caption{Projection in the $x-y$ plane of the evolution of the original cluster (\texttt{m1e4}, upper panel) and  four different generated clusters (lower panels) as a function of time. The clusters are shown at their initial configuration (first column) and at three different time steps: 1 Myr (second column), 5 Myr (third column) and 10 Myr (last column). The colour code marks the different masses of the sink particles and their generations. \label{new_generations_positions}}
\end{figure*}

Due to the redistribution of particle positions, velocities and masses in the generation process, the value of the total virial ratio $\alpha_{\rm vir}$ may be significantly altered (with respect, in this case, to the value of $1.19$ for the \texttt{m1e4} case) ranging from a minimum of $0.46$ to a maximum of $2.08$. Clearly, the future dynamical evolution depends heavily on the virial ratio, which, in turn, is heavily affected by the left tail of the particle pairwise distances. There is indeed a margin of variation in short distances between realizations, as shown in Fig.~\ref{fd}. However, the shortest distances in any stellar system essentially correspond to binary-star semiaxes. 
Our hydro-dynamical simulations were not designed to faithfully reproduce an observational initial mass function \citep[][]{2021MNRAS.501.2920B} nor to capture binary properties. In \cite{torniamenti21}, we introduce a realistic binary distribution with a separate procedure. While binary binding energy is a large fraction of the total binding energy in many realistic scenarios, the time scale over which this energy is exchanged with the cluster at large is much longer than the dissolution time for the typical system under consideration: hard binaries are dynamically inert in the short term.
To check that this is the source of the observed virial ratio mismatch, we have operated two diagnostics. First of all, we have recomputed the virial ratio $\alpha_{\rm vir}$ for $N_s$ times excluding each time a different particle. This gives us a robust way to quantify the virial ratio, as the spread in the resulting distribution will be driven by instances in which a member of a very close binary was excluded. In all cases, the value of $\alpha_{\rm vir}$ of the original system lies well inside the distribution of $\alpha_{\rm vir}$ obtained by removing one particle at time from a given generation.

Second, we have also computed the $\alpha_{\rm vir}$ by excluding the binding energy of stars with separation under a varying threshold between one tenth and one half of the average inter-particle distance. We found that the large variations in the value of $\alpha_{\rm vir}$ observed for the generated clusters is essentially due to the different distributions of tightly bound particles in the generated clusters and the parent sink particle system produced by our SPH simulations. Thus, different values of the virial ratio will result in a similar dynamical evolution on the time scales of interest, as shown below by evolving our realization through direct $N-$body simulations. We list the nominal virial coefficients of our generated realizations together with other properties in Table~\ref{table2}. 

\subsection{\textit{N}-body simulations}
Our method aims to generate large samples of initial conditions for $N-$body simulations.  To test that our realizations are indeed suitable for this use, we evolve via direct $N$-body simulations the three original clusters (\texttt{m1e4}, \texttt{m3e4} and \texttt{m6e4}) and 10 different generated clusters per each of the three original ones.  Finding that the evolution of the generated clusters is neither identical nor dramatically  different with respect to the original cluster is one of the main test-beds of our method. In fact, our method can be successfully used only if the new clusters evolve in a similar way as the original one, but are sufficiently different not to be an exact copy. Ideally, the generated clusters should behave as different random realizations of the same underlying physical distributions.

\begin{table}
\caption{Properties of the generated clusters starting from \texttt{m1e4}, \texttt{m3e4} and \texttt{m6e4}.}
\begin{center}
\begin{tabular}{llllll}
\hline
Name & $N_{\rm s}$ & $N_{\rm c}$ & $\alpha_{\rm vir}$ & $\gamma$ & $M_{\rm sink}$ $\left[{\rm M}_\odot\right]$  \\
\hline 
\texttt{m1e4g1} & 2006 & $6$ & $0.60$ & $2.3$ & $4.20\times 10^3$  \\
\texttt{m1e4g2} & 2509 & $6$ & $1.41$ & $2.3$ & $4.20\times 10^3$  \\
\texttt{m1e4g3} & 2512 & $6$ & $1.57$ & $2.3$ & $4.20\times 10^3$  \\
\texttt{m1e4g4} & 1998 & $8$ & $0.60$ & $2.3$ & $4.20\times 10^3$  \\
\texttt{m1e4g5} & 2512 & $5$ & $1.68$ & $2.3$ & $4.20\times 10^3$  \\
\texttt{m1e4g6} & 2491 & $7$ & $1.50$ & $2.3$ & $4.20\times 10^3$  \\
\texttt{m1e4g7} & 2081 & $8$ & $0.48$ & $2.3$ & $4.20\times 10^3$  \\
\texttt{m1e4g8} & 2512 & $9$ & $1.81$ & $2.3$ & $4.20\times 10^3$  \\
\texttt{m1e4g9} & 2196 & $4$ & $0.46$ & $2.3$ & $4.20\times 10^3$  \\
\texttt{m1e4g10} & 2496 & $7$ & $1.57$ & $2.3$ & $4.20\times 10^3$  \vspace{0.3cm}\\
\texttt{m3e4g1}  & 2765 & $5$ & $0.80$ & $2.2$ & $1.03\times 10^4$  \\
\texttt{m3e4g2}  & 2805 & $7$ & $1.39$ & $2.2$ & $1.03\times 10^4$  \\
\texttt{m3e4g3}  & 2811 & $5$ & $1.16$ & $2.2$ & $1.03\times 10^4$  \\
\texttt{m3e4g4}  & 2719 & $5$ & $1.20$ & $2.2$ & $1.03\times 10^4$  \\
\texttt{m3e4g5}  & 2747 & $7$ & $1.48$ & $2.2$ & $1.03\times 10^4$  \\
\texttt{m3e4g6}  & 2774 & $6$ & $1.40$ & $2.2$ & $1.03\times 10^4$  \\
\texttt{m3e4g7}  & 2750 & $6$ & $1.46$ & $2.2$ & $1.03\times 10^4$  \\
\texttt{m3e4g8}  & 2770 & $7$ & $1.13$ & $2.2$ & $1.03\times 10^4$  \\
\texttt{m3e4g9}  & 2628 & $7$ & $0.94$ & $2.2$ & $1.03\times 10^4$  \\
\texttt{m3e4g10} & 2764 & $6$ & $0.94$ & $2.2$ & $1.03\times 10^4$   \vspace{0.3cm}\\
\texttt{m6e4g1}  & 2747 & $7$ & $1.65$ & $2.1$ & $2.04\times 10^4$  \\
\texttt{m6e4g2}  & 2823 & $7$ & $1.80$ & $2.1$ & $2.04\times 10^4$ \\
\texttt{m6e4g3}  & 2900 & $5$ & $1.82$ & $2.1$ & $2.04\times 10^4$ \\
\texttt{m6e4g4}  & 2718 & $6$ & $1.66$ & $2.1$ & $2.04\times 10^4$ \\
\texttt{m6e4g5}  & 2967 & $6$ & $1.75$ & $2.1$ &  $2.04\times 10^4$ \\
\texttt{m6e4g6}  & 2752 & $5$ & $1.30$ & $2.1$ &  $2.04\times 10^4$ \\
\texttt{m6e4g7}  & 2998 & $6$ & $1.55$ & $2.1$ &  $2.04\times 10^4$ \\
\texttt{m6e4g8}  & 2833 & $6$ & $1.36$ & $2.1$ &  $2.04\times 10^4$ \\
\texttt{m6e4g9}  & 3001 & $6$ & $1.82$ & $2.1$ &  $2.04\times 10^4$ \\
\texttt{m6e4g10} & 3015 & $5$ & $1.82$ & $2.1$ & $2.04\times 10^4$  \\
\hline
\end{tabular}
\end{center}
\footnotesize{After the name of the generated cluster (Col.~1), we report the total number of stars (Col.~2), the number of macroscopic sub-clumps (Col.~3), the virial ratio (Col.~4), the $\gamma$ coefficient of the mass-spectrum fitting function (Eq.~ \ref{tinypebbleshugeboulders}, Col.~5), and the total mass of the stars (Col.~6).}
\label{table2}
\end{table}

We ran our simulations with the direct $N$-body code {\sc nbody6++gpu} \citep{wang15}. Thanks to a neighbour scheme \citep{nitadori12}, {\sc nbody6++gpu} efficiently handles the collisional force contributions at short time scales as well as those at longer time intervals, to which all the members in the system contribute. The force integration also includes a solar neighbourhood-like static external tidal field \citep{wang16}. Stellar evolution is not included in our runs, for the sake of simplicity and to make the comparison with the original cluster more straightforward. We evolved the clusters for $10 \; \mathrm{Myr}$.

Table~\ref{table2} shows the main initial properties of the generated clusters for which we ran the \textit{N}$-$body simulations. Figure \ref{new_generations_positions} shows the projection in the $x-y$ plane of the original \texttt{m1e4} cluster and of four generations, at different times. The global evolution of the new generated clusters shows a variety of configurations depending on the different distribution of mass. In some cases, distinct sub-clumps are present at $t > 1 \, \mathrm{Myr}$ and tidally interact with each other before eventually merging. In the case of \texttt{m1e4g4}, two distinct sub-clumps are still present at $10 \, \mathrm{Myr}$.  

\begin{figure}
	\includegraphics[width=\hsize]{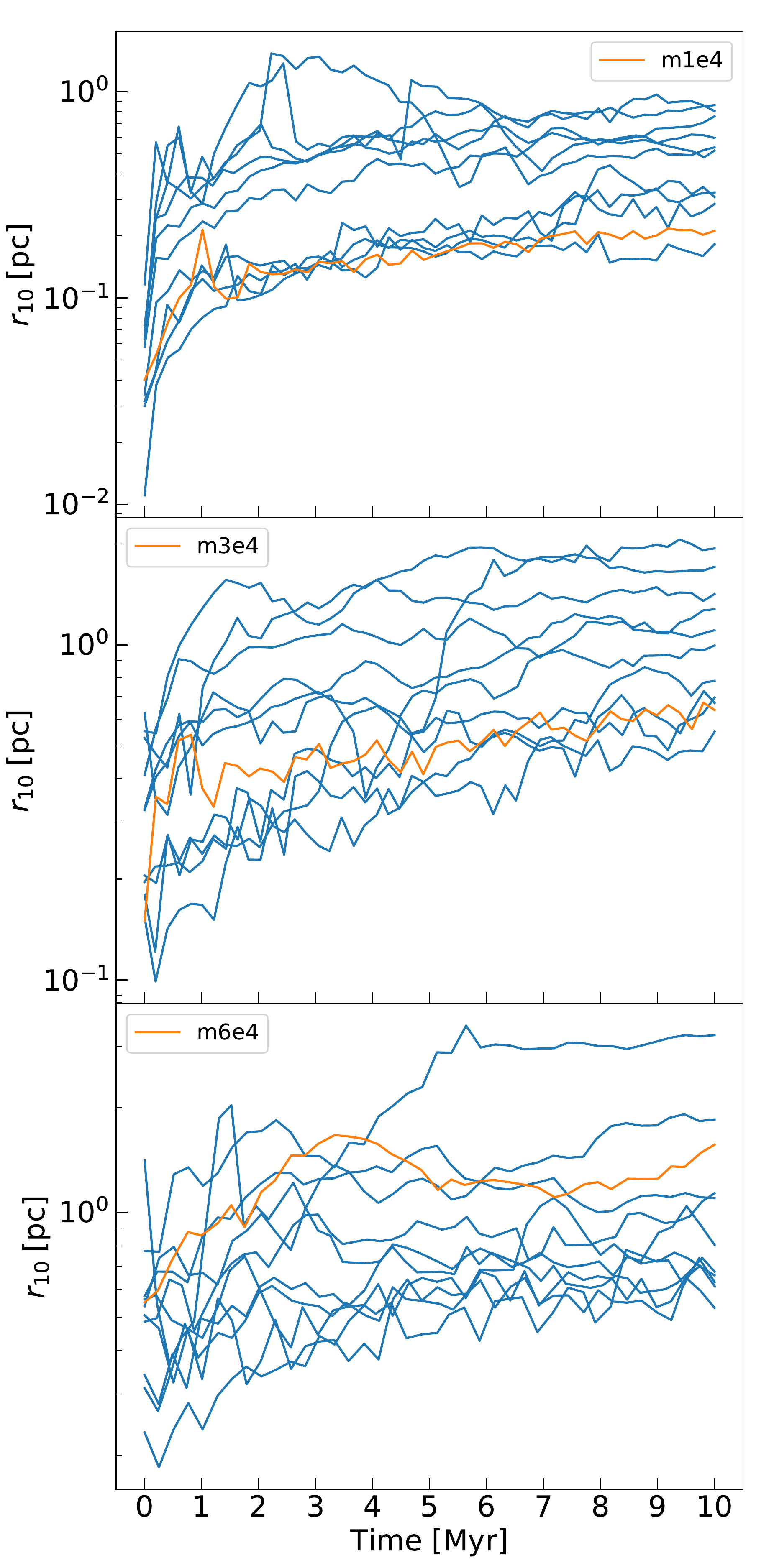}
    \caption{Evolution of the 10\% Lagrangian radius for the original sink particles and for ten different generations of \texttt{m1e4} (upper panel), \texttt{m3e4} (middle panel), and \texttt{m6e4} (lower panel). The orange line represents the original sink particle system, and the blue lines are the generated clusters. \label{new_generations_rcore}}
\end{figure}

\begin{figure}
	\includegraphics[width=\hsize]{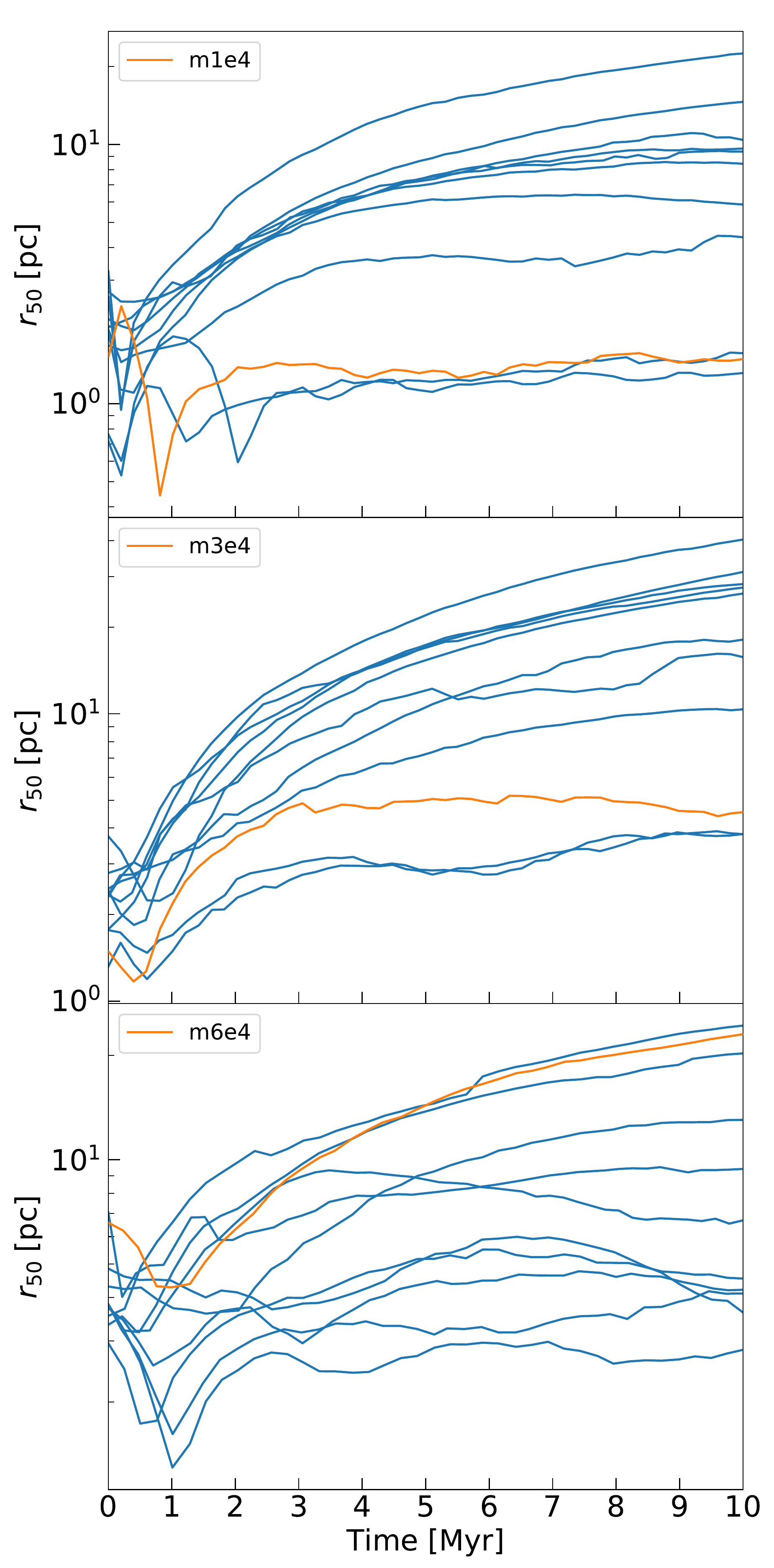}
    \caption{Same as Figure~\ref{new_generations_rcore} but for the 50\% Lagrangian radius. }\label{new_generations_rhm}
\end{figure}

A more quantitative description of the global evolution of the clusters can be given in terms of the evolution of the 10\% and 50\% Lagrangian radii ($r_\mathrm{10}$ and $r_{\mathrm{50}}$), centred in the centre of density\footnote{ The local density around each star was calculated as the density of the sphere that includes the 300 closest stars.}.  Figures \ref{new_generations_rcore} and \ref{new_generations_rhm}  show the evolution of $r_{\rm 10}$ and $r_{\mathrm{50}}$ for the original clusters and the generated ones.  
In all the cases, the original evolution lies within the limits of the distribution of the generated clusters, which shows a large spread. This spread is consistent with the large stochastic fluctuations that we expect in the evolution of such low-mass clusters \citep[see, e.g.,][]{torniamenti21}. 

\section{Discussion and Conclusions}
\label{disc}

We introduced a new method for generating a number of new realizations from a given set of initial conditions (particle masses, positions and velocities) produced by hydro-dynamical simulations. The realizations are built to display a different large scale structure, but share similar properties at smaller scales, preserving in particular the fractal dimension of the original simulation. We have shown that they can be used as initial conditions for $N-$body simulations, producing a comparable evolution to the original cluster. This suggests that our method is suitable for drawing the initial conditions of a large set of $N-$ body simulations at an infinitesimal fraction of the computational cost of generating initial conditions from a hydro-dynamical simulation. 

Our novel approach relies on informing a hierarchical clustering structure (represented as a tree) from the original initial condition data through agglomerative clustering. This is later turned into new realizations by modifying the initial branches of the tree (encoding the relations between the biggest sub-clumps in the simulation). This results in realizations with different macroscopic properties from the original one (e.g., the number of big clumps and their distances), while approximately preserving the characteristics of small scale structure responsible for most of dynamical evolution (e.g., the distribution of pairwise distances between individual stars). In principle, this scheme is very flexible, allowing to choose how much of the large scale structure we control directly, by choosing the number of initial branches we modify.

The realizations we obtained with our method qualitatively resemble the original simulation when visualized in three-dimensional space. In our case, the original distribution of stars was generated by hydro-dynamical simulations of embedded clusters, so our new realizations appear qualitatively indistinguishable from the output of these simulations.
The mass spectrum and the velocity distribution are also very similar to the original simulation. The distribution of the number of neighbours as a function of distance reveals that the fractal dimension of our realizations and that of the original simulation match on different scales (they both show a similar complex fractal pattern).

Finally, we ran direct $N-$body simulations of a sample of generated initial conditions for three different original star clusters. In all the cases, the new generations show a realistic evolution on all scales, bracketing that of the original one, as shown by the trend of the 10\% and 50\%  Lagrangian radii. 
Our analysis suggests that this method is a promising way to generate new mass and phase-space distributions from existing hydro-dynamical simulations, thus increasing our sample of initial conditions for \textit{N}$-$body simulations. The speedup in computation obtained by our new method is tremendous: generating initial conditions from hydro-dynamical simulations requires about $1.5 \times 10^5$ core hours per simulation, while our procedure takes about 10 core seconds to train the initial tree distribution and generate a new realization.


\section*{Acknowledgements}
We thank the anonymous referee for their useful comments, which helped to improve this work.
This project has received funding from the European Unions Horizon 2020 research and innovation programme under the Marie Skłodowska-Curie grant agreement No. 896248. MP's initial contribution to this material is based upon work supported by Tamkeen under the NYU Abu Dhabi Research Institute grant CAP$^3$. MM, AB and GI acknowledge financial support from the European Research Council for the ERC Consolidator grant DEMOBLACK, under contract no. 770017. PFDC acknowledges financial support from MIUR-PRIN2017 project \textit{Coarse-grained description for non-equilibrium systems and transport phenomena (CO-NEST)} n.201798CZL. MM and MCA acknowledge financial support from the Austrian National Science Foundation through FWF stand-alone grant P31154-N27. 
\section*{Data availability}
The data underlying this article will be shared on reasonable request to the corresponding authors.
\bibliographystyle{mnras}
\bibliography{manuscript.bib} 

\begin{thebibliography}{}
\makeatletter
\relax
\def\mn@urlcharsother{\let\do\@makeother \do\$\do\&\do\#\do\^\do\_\do\%\do\~}
\def\mn@doi{\begingroup\mn@urlcharsother \@ifnextchar [ {\mn@doi@}
  {\mn@doi@[]}}
\def\mn@doi@[#1]#2{\def\@tempa{#1}\ifx\@tempa\@empty \href
  {http://dx.doi.org/#2} {doi:#2}\else \href {http://dx.doi.org/#2} {#1}\fi
  \endgroup}
\def\mn@eprint#1#2{\mn@eprint@#1:#2::\@nil}
\def\mn@eprint@arXiv#1{\href {http://arxiv.org/abs/#1} {{\tt arXiv:#1}}}
\def\mn@eprint@dblp#1{\href {http://dblp.uni-trier.de/rec/bibtex/#1.xml}
  {dblp:#1}}
\def\mn@eprint@#1:#2:#3:#4\@nil{\def\@tempa {#1}\def\@tempb {#2}\def\@tempc
  {#3}\ifx \@tempc \@empty \let \@tempc \@tempb \let \@tempb \@tempa \fi \ifx
  \@tempb \@empty \def\@tempb {arXiv}\fi \@ifundefined
  {mn@eprint@\@tempb}{\@tempb:\@tempc}{\expandafter \expandafter \csname
  mn@eprint@\@tempb\endcsname \expandafter{\@tempc}}}

\bibitem[\protect\citeauthoryear{{Allison}, {Goodwin}, {Parker}, {Portegies
  Zwart}  \& {de Grijs}}{{Allison} et~al.}{2010}]{2010MNRAS.407.1098A}
{Allison} R.~J.,  {Goodwin} S.~P.,  {Parker} R.~J.,  {Portegies Zwart} S.~F.,
  {de Grijs} R.,  2010, \mn@doi [\mnras] {10.1111/j.1365-2966.2010.16939.x},
  \href {https://ui.adsabs.harvard.edu/abs/2010MNRAS.407.1098A} {407, 1098}

\bibitem[\protect\citeauthoryear{{An} \& {Evans}}{{An} \&
  {Evans}}{2006}]{2006AJ....131..782A}
{An} J.~H.,  {Evans} N.~W.,  2006, \mn@doi [\aj] {10.1086/499305}, \href
  {https://ui.adsabs.harvard.edu/abs/2006AJ....131..782A} {131, 782}

\bibitem[\protect\citeauthoryear{{Ballone}, {Mapelli}, {Di Carlo},
  {Torniamenti}, {Spera}  \& {Rastello}}{{Ballone}
  et~al.}{2020}]{2020MNRAS.496...49B}
{Ballone} A.,  {Mapelli} M.,  {Di Carlo} U.~N.,  {Torniamenti} S.,  {Spera} M.,
    {Rastello} S.,  2020, \mn@doi [\mnras] {10.1093/mnras/staa1383}, \href
  {https://ui.adsabs.harvard.edu/abs/2020MNRAS.496...49B} {496, 49}

\bibitem[\protect\citeauthoryear{{Ballone}, {Torniamenti}, {Mapelli}, {Di
  Carlo}, {Spera}, {Rastello}, {Gaspari}  \& {Iorio}}{{Ballone}
  et~al.}{2021}]{2021MNRAS.501.2920B}
{Ballone} A.,  {Torniamenti} S.,  {Mapelli} M.,  {Di Carlo} U.~N.,  {Spera} M.,
   {Rastello} S.,  {Gaspari} N.,   {Iorio} G.,  2021, \mn@doi [\mnras]
  {10.1093/mnras/staa3763}, \href
  {https://ui.adsabs.harvard.edu/abs/2021MNRAS.501.2920B} {501, 2920}

\bibitem[\protect\citeauthoryear{{Bastian}, {Gieles}, {Ercolano}  \&
  {Gutermuth}}{{Bastian} et~al.}{2009}]{2009MNRAS.392..868B}
{Bastian} N.,  {Gieles} M.,  {Ercolano} B.,   {Gutermuth} R.,  2009, \mn@doi
  [\mnras] {10.1111/j.1365-2966.2008.14107.x}, \href
  {https://ui.adsabs.harvard.edu/abs/2009MNRAS.392..868B} {392, 868}

\bibitem[\protect\citeauthoryear{{Bate}}{{Bate}}{2009a}]{Bate09}
{Bate} M.~R.,  2009a, \mn@doi [\mnras] {10.1111/j.1365-2966.2008.14165.x},
  \href {https://ui.adsabs.harvard.edu/abs/2009MNRAS.392.1363B} {392, 1363}

\bibitem[\protect\citeauthoryear{{Bate}}{{Bate}}{2009b}]{bate09b}
{Bate} M.~R.,  2009b, \mn@doi [\mnras] {10.1111/j.1365-2966.2009.14970.x},
  \href {https://ui.adsabs.harvard.edu/abs/2009MNRAS.397..232B} {397, 232}

\bibitem[\protect\citeauthoryear{{Bate}, {Bonnell}  \& {Price}}{{Bate}
  et~al.}{1995}]{Bate95}
{Bate} M.~R.,  {Bonnell} I.~A.,   {Price} N.~M.,  1995, \mn@doi [\mnras]
  {10.1093/mnras/277.2.362}, \href
  {https://ui.adsabs.harvard.edu/abs/1995MNRAS.277..362B} {277, 362}

\bibitem[\protect\citeauthoryear{{Baumgardt} \& {Kroupa}}{{Baumgardt} \&
  {Kroupa}}{2007}]{Baumgardt07}
{Baumgardt} H.,  {Kroupa} P.,  2007, \mn@doi [\mnras]
  {10.1111/j.1365-2966.2007.12209.x}, \href
  {https://ui.adsabs.harvard.edu/abs/2007MNRAS.380.1589B} {380, 1589}

\bibitem[\protect\citeauthoryear{Beaumont \& Stepney}{Beaumont \&
  Stepney}{2009}]{beaumont2009grammatical}
Beaumont D.,  Stepney S.,  2009, in 2009 IEEE Congress on Evolutionary
  Computation. pp 2446--2453

\bibitem[\protect\citeauthoryear{{Bertin} \& {Stiavelli}}{{Bertin} \&
  {Stiavelli}}{1984}]{1984A&A...137...26B}
{Bertin} G.,  {Stiavelli} M.,  1984, \aap, \href
  {https://ui.adsabs.harvard.edu/abs/1984A&A...137...26B} {137, 26}

\bibitem[\protect\citeauthoryear{{Bianchini}, {Varri}, {Bertin}  \&
  {Zocchi}}{{Bianchini} et~al.}{2013}]{2013ApJ...772...67B}
{Bianchini} P.,  {Varri} A.~L.,  {Bertin} G.,   {Zocchi} A.,  2013, \mn@doi
  [\apj] {10.1088/0004-637X/772/1/67}, \href
  {https://ui.adsabs.harvard.edu/abs/2013ApJ...772...67B} {772, 67}

\bibitem[\protect\citeauthoryear{{Boekholt} \& {Portegies Zwart}}{{Boekholt} \&
  {Portegies Zwart}}{2015}]{2015ComAC...2....2B}
{Boekholt} T.,  {Portegies Zwart} S.,  2015, \mn@doi [Computational
  Astrophysics and Cosmology] {10.1186/s40668-014-0005-3}, \href
  {https://ui.adsabs.harvard.edu/abs/2015ComAC...2....2B} {2, 2}

\bibitem[\protect\citeauthoryear{{Boley}}{{Boley}}{2009}]{Boley09}
{Boley} A.~C.,  2009, \mn@doi [\apjl] {10.1088/0004-637X/695/1/L53}, \href
  {https://ui.adsabs.harvard.edu/abs/2009ApJ...695L..53B} {695, L53}

\bibitem[\protect\citeauthoryear{{Boley}, {Hayfield}, {Mayer}  \&
  {Durisen}}{{Boley} et~al.}{2010}]{Boley10}
{Boley} A.~C.,  {Hayfield} T.,  {Mayer} L.,   {Durisen} R.~H.,  2010, \mn@doi
  [\icarus] {10.1016/j.icarus.2010.01.015}, \href
  {https://ui.adsabs.harvard.edu/abs/2010Icar..207..509B} {207, 509}

\bibitem[\protect\citeauthoryear{{Bonnell}, {Bate}  \& {Vine}}{{Bonnell}
  et~al.}{2003}]{2003MNRAS.343..413B}
{Bonnell} I.~A.,  {Bate} M.~R.,   {Vine} S.~G.,  2003, \mn@doi [\mnras]
  {10.1046/j.1365-8711.2003.06687.x}, \href
  {https://ui.adsabs.harvard.edu/abs/2003MNRAS.343..413B} {343, 413}

\bibitem[\protect\citeauthoryear{{Burgers}}{{Burgers}}{1948}]{Burgers48}
{Burgers} J.~M.,  1948, \mn@doi [Advances in Applied Mechanics]
  {10.1016/S0065-2156(08)70100-5}, 1, 171

\bibitem[\protect\citeauthoryear{{Cantat-Gaudin} et~al.,}{{Cantat-Gaudin}
  et~al.}{2019}]{Cantat-Gaudin19}
{Cantat-Gaudin} T.,  et~al., 2019, \mn@doi [\aap]
  {10.1051/0004-6361/201834957}, \href
  {https://ui.adsabs.harvard.edu/abs/2019A&A...626A..17C} {626, A17}

\bibitem[\protect\citeauthoryear{{Cartwright}}{{Cartwright}}{2009}]{Cartwright09}
{Cartwright} A.,  2009, \mn@doi [\mnras] {10.1111/j.1365-2966.2009.15540.x},
  \href {https://ui.adsabs.harvard.edu/abs/2009MNRAS.400.1427C} {400, 1427}

\bibitem[\protect\citeauthoryear{Chomsky}{Chomsky}{1959}]{CHOMSKY1959137}
Chomsky N.,  1959, \mn@doi [Information and Control]
  {https://doi.org/10.1016/S0019-9958(59)90362-6}, 2, 137

\bibitem[\protect\citeauthoryear{{Claydon}, {Gieles}, {Varri}, {Heggie}  \&
  {Zocchi}}{{Claydon} et~al.}{2019}]{2019MNRAS.487..147C}
{Claydon} I.,  {Gieles} M.,  {Varri} A.~L.,  {Heggie} D.~C.,   {Zocchi} A.,
  2019, \mn@doi [\mnras] {10.1093/mnras/stz1109}, \href
  {https://ui.adsabs.harvard.edu/abs/2019MNRAS.487..147C} {487, 147}

\bibitem[\protect\citeauthoryear{{Corsaro} et~al.,}{{Corsaro}
  et~al.}{2017}]{2017NatAs...1E..64C}
{Corsaro} E.,  et~al., 2017, \mn@doi [Nature Astronomy]
  {10.1038/s41550-017-0064}, \href
  {https://ui.adsabs.harvard.edu/abs/2017NatAs...1E..64C} {1, 0064}

\bibitem[\protect\citeauthoryear{{D'Alessio}, {Calvet}  \&
  {Hartmann}}{{D'Alessio} et~al.}{2001}]{Dalessio01}
{D'Alessio} P.,  {Calvet} N.,   {Hartmann} L.,  2001, \mn@doi [\apj]
  {10.1086/320655}, \href
  {https://ui.adsabs.harvard.edu/abs/2001ApJ...553..321D} {553, 321}

\bibitem[\protect\citeauthoryear{{Dale}, {Ercolano}  \& {Bonnell}}{{Dale}
  et~al.}{2015}]{Dale15}
{Dale} J.~E.,  {Ercolano} B.,   {Bonnell} I.~A.,  2015, \mn@doi [\mnras]
  {10.1093/mnras/stv913}, \href
  {https://ui.adsabs.harvard.edu/abs/2015MNRAS.451..987D} {451, 987}

\bibitem[\protect\citeauthoryear{{Dalessandro}, {Raso}, {Kamann}, {Bellazzini},
  {Vesperini}, {Bellini}  \& {Beccari}}{{Dalessandro}
  et~al.}{2021}]{2021MNRAS.506..813D}
{Dalessandro} E.,  {Raso} S.,  {Kamann} S.,  {Bellazzini} M.,  {Vesperini} E.,
  {Bellini} A.,   {Beccari} G.,  2021, \mn@doi [\mnras]
  {10.1093/mnras/stab1257}, \href
  {https://ui.adsabs.harvard.edu/abs/2021MNRAS.506..813D} {506, 813}

\bibitem[\protect\citeauthoryear{{Daniel}, {Heggie}  \& {Varri}}{{Daniel}
  et~al.}{2017}]{2017MNRAS.468.1453D}
{Daniel} K.~J.,  {Heggie} D.~C.,   {Varri} A.~L.,  2017, \mn@doi [\mnras]
  {10.1093/mnras/stx571}, \href
  {https://ui.adsabs.harvard.edu/abs/2017MNRAS.468.1453D} {468, 1453}

\bibitem[\protect\citeauthoryear{{Davis}, {Efstathiou}, {Frenk}  \&
  {White}}{{Davis} et~al.}{1985}]{Davis1985}
{Davis} M.,  {Efstathiou} G.,  {Frenk} C.~S.,   {White} S.~D.~M.,  1985,
  \mn@doi [\apj] {10.1086/163168}, \href
  {https://ui.adsabs.harvard.edu/abs/1985ApJ...292..371D} {292, 371}

\bibitem[\protect\citeauthoryear{{Di Carlo}, {Giacobbo}, {Mapelli}, {Pasquato},
  {Spera}, {Wang}  \& {Haardt}}{{Di Carlo} et~al.}{2019}]{2019MNRAS.487.2947D}
{Di Carlo} U.~N.,  {Giacobbo} N.,  {Mapelli} M.,  {Pasquato} M.,  {Spera} M.,
  {Wang} L.,   {Haardt} F.,  2019, \mn@doi [\mnras] {10.1093/mnras/stz1453},
  \href {https://ui.adsabs.harvard.edu/abs/2019MNRAS.487.2947D} {487, 2947}

\bibitem[\protect\citeauthoryear{{Di Cintio} \& {Casetti}}{{Di Cintio} \&
  {Casetti}}{2019}]{2019MNRAS.489.5876D}
{Di Cintio} P.,  {Casetti} L.,  2019, \mn@doi [\mnras] {10.1093/mnras/stz2531},
  \href {https://ui.adsabs.harvard.edu/abs/2019MNRAS.489.5876D} {489, 5876}

\bibitem[\protect\citeauthoryear{{Di Cintio} \& {Casetti}}{{Di Cintio} \&
  {Casetti}}{2020}]{2020IAUS..351..426D}
{Di Cintio} P.,  {Casetti} L.,  2020, in {Bragaglia} A.,  {Davies} M.,  {Sills}
  A.,   {Vesperini} E.,  eds, ~ Vol. 351, Star Clusters: From the Milky Way to
  the Early Universe. pp 426--429 (\mn@eprint {arXiv} {1907.12774}),
  \mn@doi{10.1017/S1743921319006744}

\bibitem[\protect\citeauthoryear{{Dib} \& {Henning}}{{Dib} \&
  {Henning}}{2019}]{2019A&A...629A.135D}
{Dib} S.,  {Henning} T.,  2019, \mn@doi [\aap] {10.1051/0004-6361/201834080},
  \href {https://ui.adsabs.harvard.edu/abs/2019A&A...629A.135D} {629, A135}

\bibitem[\protect\citeauthoryear{{Diemand}, {Kuhlen}  \& {Madau}}{{Diemand}
  et~al.}{2006}]{Diemand2006}
{Diemand} J.,  {Kuhlen} M.,   {Madau} P.,  2006, \mn@doi [\apj]
  {10.1086/506377}, \href
  {https://ui.adsabs.harvard.edu/abs/2006ApJ...649....1D} {649, 1}

\bibitem[\protect\citeauthoryear{Eddy}{Eddy}{2004}]{eddy2004hidden}
Eddy S.~R.,  2004, Nature biotechnology, 22, 1315

\bibitem[\protect\citeauthoryear{{Elmegreen}, {Elmegreen}, {Chandar},
  {Whitmore}  \& {Regan}}{{Elmegreen} et~al.}{2006}]{elmegreen06}
{Elmegreen} B.~G.,  {Elmegreen} D.~M.,  {Chandar} R.,  {Whitmore} B.,   {Regan}
  M.,  2006, \mn@doi [\apj] {10.1086/503797}, \href
  {https://ui.adsabs.harvard.edu/abs/2006ApJ...644..879E} {644, 879}

\bibitem[\protect\citeauthoryear{{Fabricius} et~al.,}{{Fabricius}
  et~al.}{2014}]{2014ApJ...787L..26F}
{Fabricius} M.~H.,  et~al., 2014, \mn@doi [\apjl]
  {10.1088/2041-8205/787/2/L26}, \href
  {https://ui.adsabs.harvard.edu/abs/2014ApJ...787L..26F} {787, L26}

\bibitem[\protect\citeauthoryear{{Federrath}}{{Federrath}}{2013}]{federrath13}
{Federrath} C.,  2013, \mn@doi [\mnras] {10.1093/mnras/stt1644}, \href
  {https://ui.adsabs.harvard.edu/abs/2013MNRAS.436.1245F} {436, 1245}

\bibitem[\protect\citeauthoryear{{Federrath} \& {Klessen}}{{Federrath} \&
  {Klessen}}{2012}]{2012ApJ...761..156F}
{Federrath} C.,  {Klessen} R.~S.,  2012, \mn@doi [\apj]
  {10.1088/0004-637X/761/2/156}, \href
  {https://ui.adsabs.harvard.edu/abs/2012ApJ...761..156F} {761, 156}

\bibitem[\protect\citeauthoryear{{Feng} \& {Modi}}{{Feng} \&
  {Modi}}{2017}]{2017A&C....20...44F}
{Feng} Y.,  {Modi} C.,  2017, \mn@doi [Astronomy and Computing]
  {10.1016/j.ascom.2017.05.004}, \href
  {https://ui.adsabs.harvard.edu/abs/2017A&C....20...44F} {20, 44}

\bibitem[\protect\citeauthoryear{{Ferraro} et~al.,}{{Ferraro}
  et~al.}{2018}]{2018ApJ...860...50F}
{Ferraro} F.~R.,  et~al., 2018, \mn@doi [\apj] {10.3847/1538-4357/aabe2f},
  \href {https://ui.adsabs.harvard.edu/abs/2018ApJ...860...50F} {860, 50}

\bibitem[\protect\citeauthoryear{{Fujii} \& {Portegies Zwart}}{{Fujii} \&
  {Portegies Zwart}}{2016}]{2016ApJ...817....4F}
{Fujii} M.~S.,  {Portegies Zwart} S.,  2016, \mn@doi [\apj]
  {10.3847/0004-637X/817/1/4}, \href
  {https://ui.adsabs.harvard.edu/abs/2016ApJ...817....4F} {817, 4}

\bibitem[\protect\citeauthoryear{{Gavagnin}, {Bleuler}, {Rosdahl}  \&
  {Teyssier}}{{Gavagnin} et~al.}{2017}]{Gavagnin17}
{Gavagnin} E.,  {Bleuler} A.,  {Rosdahl} J.,   {Teyssier} R.,  2017, \mn@doi
  [\mnras] {10.1093/mnras/stx2222}, \href
  {https://ui.adsabs.harvard.edu/abs/2017MNRAS.472.4155G} {472, 4155}

\bibitem[\protect\citeauthoryear{{Geen}, {Hennebelle}, {Tremblin}  \&
  {Rosdahl}}{{Geen} et~al.}{2016}]{Geen16}
{Geen} S.,  {Hennebelle} P.,  {Tremblin} P.,   {Rosdahl} J.,  2016, \mn@doi
  [\mnras] {10.1093/mnras/stw2235}, \href
  {https://ui.adsabs.harvard.edu/abs/2016MNRAS.463.3129G} {463, 3129}

\bibitem[\protect\citeauthoryear{{Gieles} \& {Zocchi}}{{Gieles} \&
  {Zocchi}}{2015}]{2015MNRAS.454..576G}
{Gieles} M.,  {Zocchi} A.,  2015, \mn@doi [\mnras] {10.1093/mnras/stv1848},
  \href {https://ui.adsabs.harvard.edu/abs/2015MNRAS.454..576G} {454, 576}

\bibitem[\protect\citeauthoryear{{Goodfellow}, {Pouget-Abadie}, {Mirza}, {Xu},
  {Warde-Farley}, {Ozair}, {Courville}  \& {Bengio}}{{Goodfellow}
  et~al.}{2014}]{2014arXiv1406.2661G}
{Goodfellow} I.~J.,  {Pouget-Abadie} J.,  {Mirza} M.,  {Xu} B.,  {Warde-Farley}
  D.,  {Ozair} S.,  {Courville} A.,   {Bengio} Y.,  2014, arXiv e-prints, \href
  {https://ui.adsabs.harvard.edu/abs/2014arXiv1406.2661G} {p. arXiv:1406.2661}

\bibitem[\protect\citeauthoryear{{Goodman}, {Heggie}  \& {Hut}}{{Goodman}
  et~al.}{1993}]{1993ApJ...415..715G}
{Goodman} J.,  {Heggie} D.~C.,   {Hut} P.,  1993, \mn@doi [\apj]
  {10.1086/173196}, \href
  {https://ui.adsabs.harvard.edu/abs/1993ApJ...415..715G} {415, 715}

\bibitem[\protect\citeauthoryear{{Goodwin} \& {Bastian}}{{Goodwin} \&
  {Bastian}}{2006}]{goodwin06}
{Goodwin} S.~P.,  {Bastian} N.,  2006, \mn@doi [\mnras]
  {10.1111/j.1365-2966.2006.11078.x}, \href
  {https://ui.adsabs.harvard.edu/abs/2006MNRAS.373..752G} {373, 752}

\bibitem[\protect\citeauthoryear{{Goodwin} \& {Whitworth}}{{Goodwin} \&
  {Whitworth}}{2004}]{2004A&A...413..929G}
{Goodwin} S.~P.,  {Whitworth} A.~P.,  2004, \mn@doi [\aap]
  {10.1051/0004-6361:20031529}, \href
  {https://ui.adsabs.harvard.edu/abs/2004A&A...413..929G} {413, 929}

\bibitem[\protect\citeauthoryear{{Hemsendorf} \& {Merritt}}{{Hemsendorf} \&
  {Merritt}}{2002}]{2002ApJ...580..606H}
{Hemsendorf} M.,  {Merritt} D.,  2002, \mn@doi [\apj] {10.1086/343027}, \href
  {https://ui.adsabs.harvard.edu/abs/2002ApJ...580..606H} {580, 606}

\bibitem[\protect\citeauthoryear{{H{\'e}nault-Brunet}
  et~al.,}{{H{\'e}nault-Brunet} et~al.}{2012}]{Henault-Brunet12}
{H{\'e}nault-Brunet} V.,  et~al., 2012, \mn@doi [\aap]
  {10.1051/0004-6361/201219472}, \href
  {https://ui.adsabs.harvard.edu/abs/2012A&A...545L...1H} {545, L1}

\bibitem[\protect\citeauthoryear{{Hills}}{{Hills}}{1980}]{hills80}
{Hills} J.~G.,  1980, \mn@doi [\apj] {10.1086/157703}, \href
  {https://ui.adsabs.harvard.edu/abs/1980ApJ...235..986H} {235, 986}

\bibitem[\protect\citeauthoryear{Jelinek, Lafferty  \& Mercer}{Jelinek
  et~al.}{1992}]{10.1007/978-3-642-76626-8_35}
Jelinek F.,  Lafferty J.~D.,   Mercer R.~L.,  1992, in Laface P.,  De~Mori R.,
  eds, Speech Recognition and Understanding. Springer Berlin Heidelberg,
  Berlin, Heidelberg, pp 345--360

\bibitem[\protect\citeauthoryear{{Kamann} et~al.,}{{Kamann}
  et~al.}{2018}]{2018MNRAS.473.5591K}
{Kamann} S.,  et~al., 2018, \mn@doi [\mnras] {10.1093/mnras/stx2719}, \href
  {https://ui.adsabs.harvard.edu/abs/2018MNRAS.473.5591K} {473, 5591}

\bibitem[\protect\citeauthoryear{{Kandrup} \& {Sideris}}{{Kandrup} \&
  {Sideris}}{2003}]{2003ApJ...585..244K}
{Kandrup} H.~E.,  {Sideris} I.~V.,  2003, \mn@doi [\apj] {10.1086/345948},
  \href {https://ui.adsabs.harvard.edu/abs/2003ApJ...585..244K} {585, 244}

\bibitem[\protect\citeauthoryear{{Kaufman} \& {Rousseeuw}}{{Kaufman} \&
  {Rousseeuw}}{1990}]{1990fgda.book.....K}
{Kaufman} L.,  {Rousseeuw} P.~J.,  1990, {Finding groups in data. an
  introduction to cluster analysis}.
Wiley Series in Probability and Statistics

\bibitem[\protect\citeauthoryear{{King}}{{King}}{1966}]{1966AJ.....71...64K}
{King} I.~R.,  1966, \mn@doi [\aj] {10.1086/109857}, \href
  {https://ui.adsabs.harvard.edu/abs/1966AJ.....71...64K} {71, 64}

\bibitem[\protect\citeauthoryear{{Klessen} \& {Burkert}}{{Klessen} \&
  {Burkert}}{2000}]{2000ApJS..128..287K}
{Klessen} R.~S.,  {Burkert} A.,  2000, \mn@doi [\apjs] {10.1086/313371}, \href
  {https://ui.adsabs.harvard.edu/abs/2000ApJS..128..287K} {128, 287}

\bibitem[\protect\citeauthoryear{{Kolmogorov}}{{Kolmogorov}}{1941}]{kolmogorov41}
{Kolmogorov} A.,  1941, Akademiia Nauk SSSR Doklady, \href
  {https://ui.adsabs.harvard.edu/abs/1941DoSSR..30..301K} {30, 301}

\bibitem[\protect\citeauthoryear{{Krumholz}, {Klein}  \& {McKee}}{{Krumholz}
  et~al.}{2012}]{Krumholz12}
{Krumholz} M.~R.,  {Klein} R.~I.,   {McKee} C.~F.,  2012, \mn@doi [\apj]
  {10.1088/0004-637X/754/1/71}, \href
  {https://ui.adsabs.harvard.edu/abs/2012ApJ...754...71K} {754, 71}

\bibitem[\protect\citeauthoryear{{Kuhn}, {Hillenbrand}, {Sills}, {Feigelson}
  \& {Getman}}{{Kuhn} et~al.}{2019}]{Kuhn19}
{Kuhn} M.~A.,  {Hillenbrand} L.~A.,  {Sills} A.,  {Feigelson} E.~D.,   {Getman}
  K.~V.,  2019, \mn@doi [\apj] {10.3847/1538-4357/aaef8c}, \href
  {https://ui.adsabs.harvard.edu/abs/2019ApJ...870...32K} {870, 32}

\bibitem[\protect\citeauthoryear{{K{\"u}pper}, {Maschberger}, {Kroupa}  \&
  {Baumgardt}}{{K{\"u}pper} et~al.}{2011a}]{2011ascl.soft07015K}
{K{\"u}pper} A.~H.~W.,  {Maschberger} T.,  {Kroupa} P.,   {Baumgardt} H.,
  2011a, {McLuster: A Tool to Make a Star Cluster} (\mn@eprint {ascl}
  {1107.015})

\bibitem[\protect\citeauthoryear{{K{\"u}pper}, {Maschberger}, {Kroupa}  \&
  {Baumgardt}}{{K{\"u}pper} et~al.}{2011b}]{2011MNRAS.417.2300K}
{K{\"u}pper} A. H.~W.,  {Maschberger} T.,  {Kroupa} P.,   {Baumgardt} H.,
  2011b, \mn@doi [\mnras] {10.1111/j.1365-2966.2011.19412.x}, \href
  {https://ui.adsabs.harvard.edu/abs/2011MNRAS.417.2300K} {417, 2300}

\bibitem[\protect\citeauthoryear{{Lada} \& {Lada}}{{Lada} \&
  {Lada}}{2003}]{2003ARA&A..41...57L}
{Lada} C.~J.,  {Lada} E.~A.,  2003, \mn@doi [\araa]
  {10.1146/annurev.astro.41.011802.094844}, \href
  {https://ui.adsabs.harvard.edu/abs/2003ARA&A..41...57L} {41, 57}

\bibitem[\protect\citeauthoryear{{Larson}}{{Larson}}{1995}]{larson95}
{Larson} R.~B.,  1995, \mn@doi [\mnras] {10.1093/mnras/272.1.213}, \href
  {https://ui.adsabs.harvard.edu/abs/1995MNRAS.272..213L} {272, 213}

\bibitem[\protect\citeauthoryear{{Lee} \& {Hennebelle}}{{Lee} \&
  {Hennebelle}}{2016}]{2016A&A...591A..30L}
{Lee} Y.-N.,  {Hennebelle} P.,  2016, \mn@doi [\aap]
  {10.1051/0004-6361/201527981}, \href
  {https://ui.adsabs.harvard.edu/abs/2016A&A...591A..30L} {591, A30}

\bibitem[\protect\citeauthoryear{{Lee} \& {Hennebelle}}{{Lee} \&
  {Hennebelle}}{2019}]{Lee19}
{Lee} Y.-N.,  {Hennebelle} P.,  2019, \mn@doi [\aap]
  {10.1051/0004-6361/201834428}, \href
  {https://ui.adsabs.harvard.edu/abs/2019A&A...622A.125L} {622, A125}

\bibitem[\protect\citeauthoryear{{Li}, {Vogelsberger}, {Marinacci}  \&
  {Gnedin}}{{Li} et~al.}{2019}]{Li19}
{Li} H.,  {Vogelsberger} M.,  {Marinacci} F.,   {Gnedin} O.~Y.,  2019, \mn@doi
  [\mnras] {10.1093/mnras/stz1271}, \href
  {https://ui.adsabs.harvard.edu/abs/2019MNRAS.487..364L} {487, 364}

\bibitem[\protect\citeauthoryear{Lindenmayer}{Lindenmayer}{1968a}]{LINDENMAYER1968280}
Lindenmayer A.,  1968a, \mn@doi [Journal of Theoretical Biology]
  {https://doi.org/10.1016/0022-5193(68)90079-9}, 18, 280

\bibitem[\protect\citeauthoryear{Lindenmayer}{Lindenmayer}{1968b}]{LINDENMAYER1968300}
Lindenmayer A.,  1968b, \mn@doi [Journal of Theoretical Biology]
  {https://doi.org/10.1016/0022-5193(68)90080-5}, 18, 300

\bibitem[\protect\citeauthoryear{{Lupton} \& {Gunn}}{{Lupton} \&
  {Gunn}}{1987}]{1987AJ.....93.1106L}
{Lupton} R.~H.,  {Gunn} J.~E.,  1987, \mn@doi [\aj] {10.1086/114394}, \href
  {https://ui.adsabs.harvard.edu/abs/1987AJ.....93.1106L} {93, 1106}

\bibitem[\protect\citeauthoryear{{Lynden-Bell}}{{Lynden-Bell}}{1962}]{1962MNRAS.123..447L}
{Lynden-Bell} D.,  1962, \mn@doi [\mnras] {10.1093/mnras/123.5.447}, \href
  {https://ui.adsabs.harvard.edu/abs/1962MNRAS.123..447L} {123, 447}

\bibitem[\protect\citeauthoryear{{Maciejewski}, {Colombi}, {Springel}, {Alard}
  \& {Bouchet}}{{Maciejewski} et~al.}{2009}]{Maciejewski2009}
{Maciejewski} M.,  {Colombi} S.,  {Springel} V.,  {Alard} C.,   {Bouchet}
  F.~R.,  2009, \mn@doi [\mnras] {10.1111/j.1365-2966.2009.14825.x}, \href
  {https://ui.adsabs.harvard.edu/abs/2009MNRAS.396.1329M} {396, 1329}

\bibitem[\protect\citeauthoryear{{Manwadkar}, {Trani}  \& {Leigh}}{{Manwadkar}
  et~al.}{2020}]{2020MNRAS.497.3694M}
{Manwadkar} V.,  {Trani} A.~A.,   {Leigh} N. W.~C.,  2020, \mn@doi [\mnras]
  {10.1093/mnras/staa1722}, \href
  {https://ui.adsabs.harvard.edu/abs/2020MNRAS.497.3694M} {497, 3694}

\bibitem[\protect\citeauthoryear{{Mapelli}}{{Mapelli}}{2017}]{2017MNRAS.467.3255M}
{Mapelli} M.,  2017, \mn@doi [\mnras] {10.1093/mnras/stx304}, \href
  {https://ui.adsabs.harvard.edu/abs/2017MNRAS.467.3255M} {467, 3255}

\bibitem[\protect\citeauthoryear{{Michie} \& {Bodenheimer}}{{Michie} \&
  {Bodenheimer}}{1963}]{1963MNRAS.126..269M}
{Michie} R.~W.,  {Bodenheimer} P.~H.,  1963, \mn@doi [\mnras]
  {10.1093/mnras/126.3.269}, \href
  {https://ui.adsabs.harvard.edu/abs/1963MNRAS.126..269M} {126, 269}

\bibitem[\protect\citeauthoryear{{Murphy}, {Geach}  \& {Bower}}{{Murphy}
  et~al.}{2012}]{2012MNRAS.420.1861M}
{Murphy} D.~N.~A.,  {Geach} J.~E.,   {Bower} R.~G.,  2012, \mn@doi [\mnras]
  {10.1111/j.1365-2966.2011.19782.x}, \href
  {https://ui.adsabs.harvard.edu/abs/2012MNRAS.420.1861M} {420, 1861}

\bibitem[\protect\citeauthoryear{{Nitadori} \& {Aarseth}}{{Nitadori} \&
  {Aarseth}}{2012}]{nitadori12}
{Nitadori} K.,  {Aarseth} S.~J.,  2012, \mn@doi [\mnras]
  {10.1111/j.1365-2966.2012.21227.x}, \href
  {https://ui.adsabs.harvard.edu/abs/2012MNRAS.424..545N} {424, 545}

\bibitem[\protect\citeauthoryear{{Park}, {Goodwin}  \& {Kim}}{{Park}
  et~al.}{2018}]{2018MNRAS.478..183P}
{Park} S.-M.,  {Goodwin} S.~P.,   {Kim} S.~S.,  2018, \mn@doi [\mnras]
  {10.1093/mnras/sty1083}, \href
  {https://ui.adsabs.harvard.edu/abs/2018MNRAS.478..183P} {478, 183}

\bibitem[\protect\citeauthoryear{{Parker}, {Goodwin}  \& {Allison}}{{Parker}
  et~al.}{2011}]{2011MNRAS.418.2565P}
{Parker} R.~J.,  {Goodwin} S.~P.,   {Allison} R.~J.,  2011, \mn@doi [\mnras]
  {10.1111/j.1365-2966.2011.19646.x}, \href
  {https://ui.adsabs.harvard.edu/abs/2011MNRAS.418.2565P} {418, 2565}

\bibitem[\protect\citeauthoryear{{Parker}, {Wright}, {Goodwin}  \&
  {Meyer}}{{Parker} et~al.}{2014}]{2014MNRAS.438..620P}
{Parker} R.~J.,  {Wright} N.~J.,  {Goodwin} S.~P.,   {Meyer} M.~R.,  2014,
  \mn@doi [\mnras] {10.1093/mnras/stt2231}, \href
  {https://ui.adsabs.harvard.edu/abs/2014MNRAS.438..620P} {438, 620}

\bibitem[\protect\citeauthoryear{{Pasquato} \& {Milone}}{{Pasquato} \&
  {Milone}}{2019}]{2019arXiv190604983P}
{Pasquato} M.,  {Milone} A.,  2019, arXiv e-prints, \href
  {https://ui.adsabs.harvard.edu/abs/2019arXiv190604983P} {p. arXiv:1906.04983}

\bibitem[\protect\citeauthoryear{Pedregosa et~al.,}{Pedregosa
  et~al.}{2011}]{scikit-learn}
Pedregosa F.,  et~al., 2011, Journal of Machine Learning Research, 12, 2825

\bibitem[\protect\citeauthoryear{{Pfalzner}}{{Pfalzner}}{2009}]{pfalzner09}
{Pfalzner} S.,  2009, \mn@doi [\aap] {10.1051/0004-6361/200912056}, \href
  {https://ui.adsabs.harvard.edu/abs/2009A&A...498L..37P} {498, L37}

\bibitem[\protect\citeauthoryear{{Plummer}}{{Plummer}}{1911}]{1911MNRAS..71..460P}
{Plummer} H.~C.,  1911, \mn@doi [\mnras] {10.1093/mnras/71.5.460}, \href
  {https://ui.adsabs.harvard.edu/abs/1911MNRAS..71..460P} {71, 460}

\bibitem[\protect\citeauthoryear{{Prendergast} \& {Tomer}}{{Prendergast} \&
  {Tomer}}{1970}]{1970AJ.....75..674P}
{Prendergast} K.~H.,  {Tomer} E.,  1970, \mn@doi [\aj] {10.1086/111008}, \href
  {https://ui.adsabs.harvard.edu/abs/1970AJ.....75..674P} {75, 674}

\bibitem[\protect\citeauthoryear{{Press} \& {Schechter}}{{Press} \&
  {Schechter}}{1974}]{PressSchechter1974}
{Press} W.~H.,  {Schechter} P.,  1974, \mn@doi [\apj] {10.1086/152650}, \href
  {https://ui.adsabs.harvard.edu/abs/1974ApJ...187..425P} {187, 425}

\bibitem[\protect\citeauthoryear{Prusinkiewicz \& Hanan}{Prusinkiewicz \&
  Hanan}{2013}]{prusinkiewicz2013lindenmayer}
Prusinkiewicz P.,  Hanan J.,  2013, Lindenmayer systems, fractals, and plants.
 Lecture notes in Biomathematics Vol. 79, Springer Science \& Business Media

\bibitem[\protect\citeauthoryear{Rabiner \& Juang}{Rabiner \&
  Juang}{1986}]{1165342}
Rabiner L.,  Juang B.,  1986, \mn@doi [IEEE ASSP Magazine]
  {10.1109/MASSP.1986.1165342}, 3, 4

\bibitem[\protect\citeauthoryear{{Reina-Campos}, {Kruijssen}, {Pfeffer},
  {Bastian}  \& {Crain}}{{Reina-Campos} et~al.}{2019}]{2019MNRAS.486.5838R}
{Reina-Campos} M.,  {Kruijssen} J.~M.~D.,  {Pfeffer} J.~L.,  {Bastian} N.,
  {Crain} R.~A.,  2019, \mn@doi [\mnras] {10.1093/mnras/stz1236}, \href
  {https://ui.adsabs.harvard.edu/abs/2019MNRAS.486.5838R} {486, 5838}

\bibitem[\protect\citeauthoryear{{Rodriguez-Gomez} et~al.,}{{Rodriguez-Gomez}
  et~al.}{2015}]{rodriguezgomez15}
{Rodriguez-Gomez} V.,  et~al., 2015, \mn@doi [\mnras] {10.1093/mnras/stv264},
  \href {https://ui.adsabs.harvard.edu/abs/2015MNRAS.449...49R} {449, 49}

\bibitem[\protect\citeauthoryear{{Ruthotto} \& {Haber}}{{Ruthotto} \&
  {Haber}}{2021}]{2021arXiv210305180R}
{Ruthotto} L.,  {Haber} E.,  2021, arXiv e-prints, \href
  {https://ui.adsabs.harvard.edu/abs/2021arXiv210305180R} {p. arXiv:2103.05180}

\bibitem[\protect\citeauthoryear{{Salpeter}}{{Salpeter}}{1955}]{1955ApJ...121..161S}
{Salpeter} E.~E.,  1955, \mn@doi [\apj] {10.1086/145971}, \href
  {https://ui.adsabs.harvard.edu/abs/1955ApJ...121..161S} {121, 161}

\bibitem[\protect\citeauthoryear{{Seifried} et~al.,}{{Seifried}
  et~al.}{2017}]{Seifried17}
{Seifried} D.,  et~al., 2017, \mn@doi [\mnras] {10.1093/mnras/stx2343}, \href
  {https://ui.adsabs.harvard.edu/abs/2017MNRAS.472.4797S} {472, 4797}

\bibitem[\protect\citeauthoryear{{Torniamenti}, {Ballone}, {Mapelli},
  {Gaspari}, {Di Carlo}, {Rastello}, {Giacobbo}  \& {Pasquato}}{{Torniamenti}
  et~al.}{2021}]{torniamenti21}
{Torniamenti} S.,  {Ballone} A.,  {Mapelli} M.,  {Gaspari} N.,  {Di Carlo}
  U.~N.,  {Rastello} S.,  {Giacobbo} N.,   {Pasquato} M.,  2021, \mn@doi
  [\mnras] {10.1093/mnras/stab2238}, \href
  {https://ui.adsabs.harvard.edu/abs/2021MNRAS.507.2253T} {507, 2253}

\bibitem[\protect\citeauthoryear{{Trenti} \& {Bertin}}{{Trenti} \&
  {Bertin}}{2005}]{2005A&A...429..161T}
{Trenti} M.,  {Bertin} G.,  2005, \mn@doi [\aap] {10.1051/0004-6361:20041023},
  \href {https://ui.adsabs.harvard.edu/abs/2005A&A...429..161T} {429, 161}

\bibitem[\protect\citeauthoryear{{Varri} \& {Bertin}}{{Varri} \&
  {Bertin}}{2012}]{2012A&A...540A..94V}
{Varri} A.~L.,  {Bertin} G.,  2012, \mn@doi [\aap]
  {10.1051/0004-6361/201118300}, \href
  {https://ui.adsabs.harvard.edu/abs/2012A&A...540A..94V} {540, A94}

\bibitem[\protect\citeauthoryear{{V{\'a}zquez-Semadeni}, {Col{\'\i}n},
  {G{\'o}mez}, {Ballesteros-Paredes}  \& {Watson}}{{V{\'a}zquez-Semadeni}
  et~al.}{2010}]{VazquezSemadeni10}
{V{\'a}zquez-Semadeni} E.,  {Col{\'\i}n} P.,  {G{\'o}mez} G.~C.,
  {Ballesteros-Paredes} J.,   {Watson} A.~W.,  2010, \mn@doi [\apj]
  {10.1088/0004-637X/715/2/1302}, \href
  {https://ui.adsabs.harvard.edu/abs/2010ApJ...715.1302V} {715, 1302}

\bibitem[\protect\citeauthoryear{{Wadsley}, {Stadel}  \& {Quinn}}{{Wadsley}
  et~al.}{2004}]{2004NewA....9..137W}
{Wadsley} J.~W.,  {Stadel} J.,   {Quinn} T.,  2004, \mn@doi [\na]
  {10.1016/j.newast.2003.08.004}, \href
  {https://ui.adsabs.harvard.edu/abs/2004NewA....9..137W} {9, 137}

\bibitem[\protect\citeauthoryear{{Wadsley}, {Keller}  \& {Quinn}}{{Wadsley}
  et~al.}{2017}]{2017MNRAS.471.2357W}
{Wadsley} J.~W.,  {Keller} B.~W.,   {Quinn} T.~R.,  2017, \mn@doi [\mnras]
  {10.1093/mnras/stx1643}, \href
  {https://ui.adsabs.harvard.edu/abs/2017MNRAS.471.2357W} {471, 2357}

\bibitem[\protect\citeauthoryear{{Wall}, {McMillan}, {Mac Low}, {Klessen}  \&
  {Portegies Zwart}}{{Wall} et~al.}{2019}]{Wall19}
{Wall} J.~E.,  {McMillan} S. L.~W.,  {Mac Low} M.-M.,  {Klessen} R.~S.,
  {Portegies Zwart} S.,  2019, \mn@doi [\apj] {10.3847/1538-4357/ab4db1}, \href
  {https://ui.adsabs.harvard.edu/abs/2019ApJ...887...62W} {887, 62}

\bibitem[\protect\citeauthoryear{{Wang} \& {Hernandez}}{{Wang} \&
  {Hernandez}}{2021}]{2021arXiv210410843W}
{Wang} L.,  {Hernandez} D.~M.,  2021, arXiv e-prints, \href
  {https://ui.adsabs.harvard.edu/abs/2021arXiv210410843W} {p. arXiv:2104.10843}

\bibitem[\protect\citeauthoryear{{Wang}, {Spurzem}, {Aarseth}, {Nitadori},
  {Berczik}, {Kouwenhoven}  \& {Naab}}{{Wang} et~al.}{2015}]{wang15}
{Wang} L.,  {Spurzem} R.,  {Aarseth} S.,  {Nitadori} K.,  {Berczik} P.,
  {Kouwenhoven} M.~B.~N.,   {Naab} T.,  2015, \mn@doi [\mnras]
  {10.1093/mnras/stv817}, \href
  {https://ui.adsabs.harvard.edu/abs/2015MNRAS.450.4070W} {450, 4070}

\bibitem[\protect\citeauthoryear{{Wang} et~al.,}{{Wang} et~al.}{2016}]{wang16}
{Wang} L.,  et~al., 2016, \mn@doi [\mnras] {10.1093/mnras/stw274}, \href
  {https://ui.adsabs.harvard.edu/abs/2016MNRAS.458.1450W} {458, 1450}

\bibitem[\protect\citeauthoryear{{Ward}, {Kruijssen}  \& {Rix}}{{Ward}
  et~al.}{2020}]{2020MNRAS.495..663W}
{Ward} J.~L.,  {Kruijssen} J.~M.~D.,   {Rix} H.-W.,  2020, \mn@doi [\mnras]
  {10.1093/mnras/staa1056}, \href
  {https://ui.adsabs.harvard.edu/abs/2020MNRAS.495..663W} {495, 663}

\bibitem[\protect\citeauthoryear{{Wilson}}{{Wilson}}{1975}]{1975AJ.....80..175W}
{Wilson} C.~P.,  1975, \mn@doi [\aj] {10.1086/111729}, \href
  {https://ui.adsabs.harvard.edu/abs/1975AJ.....80..175W} {80, 175}

\bibitem[\protect\citeauthoryear{{Zamora-Avil{\'e}s}
  et~al.,}{{Zamora-Avil{\'e}s} et~al.}{2019}]{Zamora-Aviles19}
{Zamora-Avil{\'e}s} M.,  et~al., 2019, \mn@doi [\mnras]
  {10.1093/mnras/stz1235}, \href
  {https://ui.adsabs.harvard.edu/abs/2019MNRAS.487.2200Z} {487, 2200}

\makeatother
\end{thebibliography}
\bsp	





\label{lastpage}
\end{document}